\RequirePackage{fix-cm}
\documentclass{svjour3}                     %
\smartqed  %
\usepackage{graphicx}
\usepackage{amsmath,amssymb,amsfonts}
\usepackage{newtxmath}
\usepackage{textcomp}
\usepackage{pgfplotstable}
\usepackage{array}
\usepackage{booktabs}
\usepackage{colortbl}
\usepackage[numbers]{natbib}
\usepackage{multirow}
\definecolor{lightgray}{rgb}{0.93, 0.93, 0.93}

\usepackage[linesnumbered, ruled, vlined]{algorithm2e}
\SetAlFnt{\small}
\SetAlCapFnt{\small}
\SetAlCapNameFnt{\small}

\usepackage[caption=false, font=footnotesize, subrefformat=parens]{subfig}

\usepackage{framed}
\usepackage{paralist}
\usepackage{xspace}
\newcommand{\app}{\emph{PRINS}\xspace}

\usepackage{hyperref}

\usepackage{datatool}
\usepackage[binary-units]{siunitx}
\DeclareSIUnit{\pp}{\textup{pp}}
\usepackage[english]{fnumprint}

\DTLloaddb{max}{summary-values.tex}

\newcommand{\minSpec}{\DTLfetch{max}{data}{all}{spec.min}}
\newcommand{\avgSpec}{\DTLfetch{max}{data}{all}{spec.avg}}
\newcommand{\maxSpec}{\DTLfetch{max}{data}{all}{spec.max}}

\newcommand{\minRecall}{\DTLfetch{max}{data}{all}{recall.min}}
\newcommand{\avgRecall}{\DTLfetch{max}{data}{all}{recall.avg}}
\newcommand{\maxRecall}{\DTLfetch{max}{data}{all}{recall.max}}

\newcommand{\avgBA}{\DTLfetch{max}{data}{all}{ba.avg}}

\def\BibTeX{{\rm B\kern-.05em{\sc i\kern-.025em b}\kern-.08em
  T\kern-.1667em\lower.7ex\hbox{E}\kern-.125emX}}

\journalname{Empirical Software Engineering}
\begin{document}
\begin{filecontents*}{summary-values.tex}
data,recall.min,recall.avg,recall.max,spec.min,spec.avg,spec.max,ba.min,ba.avg,ba.max
all,-23.5,-4.9,0,-3.1,6.3,34.9,-7,0.7,17.4
\end{filecontents*}

\title{PRINS: Scalable Model Inference for Component-based System Logs\thanks{This work has received funding from the Luxembourg National Research Fund (FNR) under grant
agreement No C-PPP17/IS/11602677 and from the NSERC Discovery and Canada Research Chair programmes.
Donghwan Shin was also partially supported by the Basic Science Research Program through
the National Research Foundation of Korea (NRF) funded by the Ministry of Education (2019R1A6A3A03033444).}}

\author{
	Donghwan Shin
	\and
    Domenico Bianculli
    \and
    Lionel Briand
}

\institute{
	Donghwan Shin \and Domenico Bianculli \and Lionel Briand \at
	University of Luxembourg \\
	\email{donghwan.shin@uni.lu, domenico.bianculli@uni.lu, lionel.briand@uni.lu}
\and
	Lionel Briand \at
	University of Ottawa \\
}

\date{Received: date / Accepted: date}

\maketitle

\begin{abstract}
Behavioral software models play a key role in many software engineering tasks;
unfortunately, these models either are not available during software development
or, if available, quickly become outdated as implementations evolve.
Model inference techniques have been proposed as a viable solution to extract
finite state models from execution logs. However, existing techniques do not
scale well when processing very large logs that can be commonly found in
practice.

In this paper, we address the scalability problem of inferring the model of a
component-based system from large system logs, without requiring any extra
information. Our model inference technique, called \app, follows a divide-and-conquer
approach. The idea is to first infer a model of each system component
from the corresponding logs; then, the individual component models are merged
together taking into account the flow of events across components, as reflected in
the logs. We evaluated \app in terms of scalability and accuracy, using nine
datasets composed of logs extracted from publicly available benchmarks and a
personal computer running desktop business applications. The results show that
\app can process large logs much faster than a publicly available and well-known
state-of-the-art tool, without significantly compromising the accuracy of
inferred models.
\end{abstract}

\section{Introduction}\label{sec:intro}
Behavior models of software system components play a key role in many software
engineering tasks, such as program comprehension~\cite{Cook:1998:287001}, test
case generation~\cite{6200086}, and model checking~\cite{clarke2018model}.
Unfortunately, such models are either scarce during software development or, if
available, quickly become outdated as the corresponding implementations evolve,
because of the time and cost involved in generating and maintaining
them~\cite{Walkinshaw2010af}.

One possible way to overcome the lack of software models is to use \emph{model
inference} techniques, which extract models---typically in the form of
Finite State Machine (FSM)---from execution logs. Although the problem
of inferring a minimal FSM is NP-complete~\cite{GOLD1967447}, there
have been several proposals of polynomial-time approximation algorithms to infer
FSMs~\cite{biermann1972synthesis,Beschastnikh2011Lev} or richer
variants, such as gFSM (guarded
FSM)~\cite{walkinshaw2016inferring,mariani2017gk} and gFSM extended with
transition probabilities~\cite{Emam2018Inf}, to obtain relatively faithful models.

Though the aforementioned model inference techniques are fast and accurate
enough for relatively small programs, all of them suffer from scalability
issues, due to the intrinsic computational complexity of the problem. This leads
to out-of-memory errors or extremely long, impractical execution time when
processing very large logs~\cite{wang2016scalable} that can be commonly found in
practice. A recent proposal~\cite{LUO201713} addresses scalability
using a distributed FSM inference approach based on MapReduce. However, this
approach requires to encode the data to be exchanged between mappers and
reducers in the form of key-value pairs. Such encoding is application-specific;
hence, it cannot be used in contexts---like the one in which this work has been
performed---in which the system is treated as a black-box (i.e., the source code
is not available), with limited information about the data recorded in the
system logs.

In this paper, we address the scalability problem of inferring a
system model from the logs recorded during the execution (possibly
multiple executions) of a system composed of multiple ``components''
(hereafter called \emph{component-based system}), without requiring
any extra information other than the logs.  In this paper, we use the
term ``component'' in a broad sense: the large majority of modern
software systems are composed of different types of ``components'',
such as modules, classes, and services; in all cases, the
resulting system decomposition provides a high degree of modularity
and separation of concerns. Our goal is to efficiently infer a system
model that captures not only the components' behaviors but also
the flow of events across the components as reflected in the logs.

Our approach, called \app, follows a \emph{divide-and-conquer}
strategy: we first infer a model of each component from the
corresponding logs using a state-of-the-art model inference technique,
and then we ``stitch'' (i.e., we do a peculiar type of merge) the
individual component models into a system-level model by taking into
account the interactions among the components, \emph{as reflected in
the logs}. The rationale behind this idea is that, though existing
model inference techniques cannot deal with the size of all combined
component logs, they can still be used to infer the models of
individual components, since their logs tend to be sufficiently small.
In other words, \app alleviates the scalability issues of existing model
inference techniques by limiting their application to the smaller scope
defined by component-level logs.

We implemented \app in a prototype tool, which internally uses
MINT~\cite{walkinshaw2016inferring}, the only publicly available
state-of-the-art technique for inferring gFSMs, to infer the
individual component models. We evaluate the scalability (in terms of
execution time) and the accuracy (in terms of recall and specificity)
of \app in comparison with MINT (to directly infer system models from
system logs), on \fnumprint{9} datasets composed of logs extracted
from publicly available benchmarks~\cite{he2020loghub} and a personal
computer (PC) running desktop business applications on a daily basis.
The results show that \app is significantly more scalable than MINT
and can even enable model inference when MINT leads to out-of-memory failures.
It also achieves higher specificity than MINT (with a
difference ranging between \SI[parse-numbers=false]{\minSpec}{\pp} and
\SI[parse-numbers=false]{+\maxSpec}{\pp}, with \si{\pp}=percentage
points) while achieving lower recall than MINT (with a difference
ranging between \SI[parse-numbers=false]{\minRecall}{\pp} and
\SI[parse-numbers=false]{+\maxRecall}{\pp}). Through a detailed
analysis, we determined that a lower recall for \app only happens when
logs are inadequate to infer accurate models, using any of the
techniques. We also propose a simple and practical metric for
engineers to easily predict (and thus improve) such cases before
running model inference. With adequate logs, \app therefore provides a
comparable or even better accuracy.

To summarize, the main contributions of this paper are:
\begin{itemize}
\item the \app approach for taming the scalability problem of
  inferring the model of a component-based system from the individual
  component-level logs, when no additional information is available;
\item the novel \emph{stitching} algorithm that ``combine'' individual
  component models together taking into account the flow of events across
  components as recorded in logs;
\item a publicly available implementation of \app (see Section\ref{sec:data});
\item the empirical evaluation, in terms of scalability and accuracy,
  of \app and its comparison with the state-of-the-art model inference tool.
\end{itemize}

The rest of the paper is organized as follows. Section~\ref{sec:background}
gives the basic definitions of logs and models that will be used throughout the
paper. Section~\ref{sec:example} illustrates the motivating example.
Section~\ref{sec:technique} describes the different stages of \app.
Section~\ref{sec:eval} reports on the evaluation of \app.
Section~\ref{sec:related} discusses related work. Section~\ref{sec:conclusion}
concludes the paper and provides directions for future work.

\section{Background}\label{sec:background}

This section provides the basic definitions for the main concepts
that will be used throughout the paper.

\subsection{Logs}\label{sec:logs}

A \emph{log} is a sequence of log entries; a \emph{log entry} contains a
timestamp (recording the time at which the logged event occurred), a component
(representing the name of the component where the event occurred), and a log
message (with run-time information related to the logged event). A log message
is typically a block of free-form text that can be further
decomposed~\cite{Zhu2019Tools,8029742,messaoudi2018search,el2020systematic} into
an event template, characterizing the event type, and the parameter values of
the event, which are determined at run time. For example, given the log entry
``\texttt{15:37:56 - Master - end (status=ok)}'', we can see that the event
\texttt{end} of the component \texttt{Master} occurred at  timestamp
\texttt{15:37:56} with the value \texttt{ok} for parameter \texttt{status}.

More formally, let $C$ be the set of all components of a system, $\mathit{ET}$
be the set of all events that can occur in the system, $V$ be the set of all
mappings from event parameters to their concrete values, for all events
$\mathit{et}\in \mathit{ET}$, and $L$ be the set of all logs retrieved for the
system; a log $l \in L$ is a sequence of log entries $\langle e_1, e_2, \dots,
e_n \rangle$, with $e_i = (\mathit{ts}_x, \mathit{cm}_i, \mathit{et}_i, v_i)$,
$\mathit{ts}_i\in \mathbb{N}$, $\mathit{cm}_i\in C$, $\mathit{et}_i\in
\mathit{ET}$, and $v_i$ is a vector of parameter values over $V$.
To denote individual log entries, we use the notation $e_{i,j}$ for
the $i$-th log entry in the $j$-th execution log.

\subsection{Models}\label{sec:gfsm}

In this paper, we represent the models inferred for a system for a component as
guarded Finite State Machines (gFSMs). Informally, a gFSM is an ``extended''
finite state machine whose transitions are triggered by the occurrence of an
event and are guarded by a function that evaluates the values of the event
parameters.

More formally, let $\mathit{ET}$ and $V$ be defined as above. A gFSM is a tuple
$m=(S, \mathit{ET}, G, \delta, s_0, F)$, where $S$ is a finite set of states,
$G$ is a finite set of guard functions of the form $g\colon V\to
\{\textit{True}, \textit{False}\}$,  $\delta$ is the transition relation $\delta
\subseteq S\times \mathit{ET} \times G \times S$, $s_0\in S$ is the initial
state, and $F\subseteq S$ is the set of final states.
A gFSM $m$ makes a guarded transition from a (source) state $s\in S$ to a
(target) state $s^\prime\in S$ when reading an input log entry $e=(\mathit{ts},
\mathit{cm}, \mathit{et}, v)$, written as $s \xrightarrow{e} s^\prime$, if
$(s,\mathit{et}, g, s^\prime) \in \delta $ and $g(v)=\textit{True}$.
We say that a guarded transition is \emph{deterministic} if there is at most one
target state for the same source state and the same log entry. Otherwise, it is
\emph{non-deterministic}. Based on this, we say that a gFSM is deterministic if
all of its guarded transitions are deterministic; otherwise, the gFSM is
non-deterministic.
We say that a gFSM $m$ \emph{accepts} a log $l = \langle e_1, \dots, e_n
\rangle$ if there exists a sequence of states $\langle \gamma_0, \dots, \gamma_n
\rangle$ such that
\begin{inparaenum}[(1)]
\item $\gamma_i \in S$ for $i=0,\dots,n$,
\item $\gamma_0 = s_0$,
\item $\gamma_{i-1} \xrightarrow{e_i} \gamma_{i}$ for $i=1,\dots,n$, and
\item $\gamma_n \in F$.
\end{inparaenum}

\section{Motivating Example}\label{sec:example}

This section presents a simple example to motivate and demonstrate our work.

Let us consider an imaginary system composed of two components,
\texttt{Master} and \texttt{Job}; \figurename~\ref{fig:example-log}
depicts the set of logs $L_\texttt{S} = \{l_1, l_2\}$ recorded during
the executions of the system. Entries in the logs are denoted using
the notation introduced in Section~\ref{sec:logs}; for instance, log
entry $e_{8,1}$ corresponds to the tuple
$(\cdot, \texttt{Master}, \textit{end}, [\textit{ok}])$, where the event
is ``\textit{end}'' and the value for its first (and only) parameter
is ``\textit{ok}''. Notice that in \figurename~\ref{fig:example-log}
we use a short form (as in ``\textit{end (ok)}'') to indicate both an
event and its parameter value; also, we omit timestamps in the
running example logs as they are not used in our approach.

\begin{figure}
	\centering
	\subfloat[$m_\texttt{S}$\label{fig:system}]{\includegraphics[height=4.5cm]{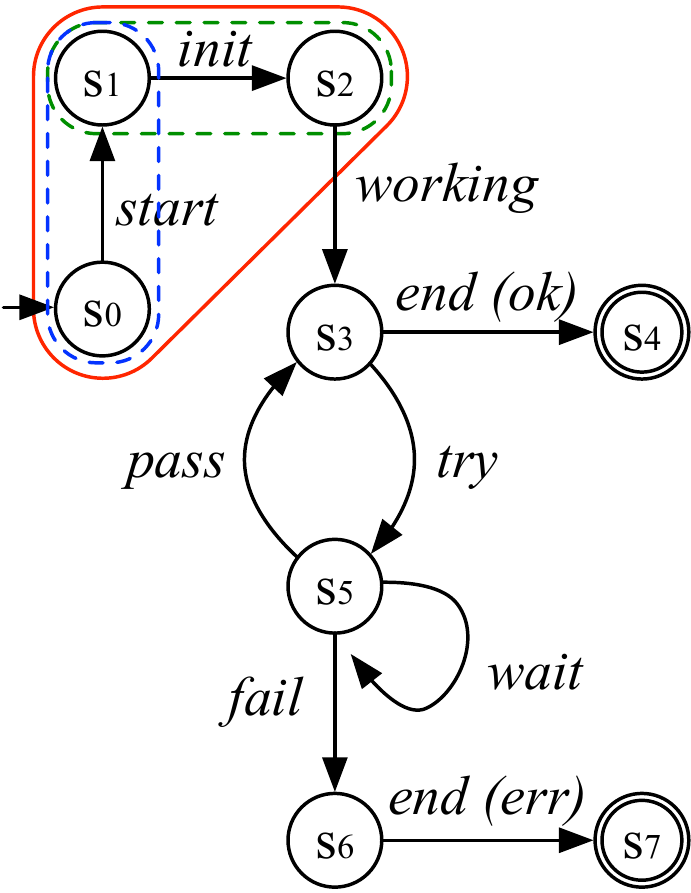}}
	\hfill
	\subfloat[$m_\texttt{M}$\label{fig:master}]{\includegraphics[height=4.5cm]{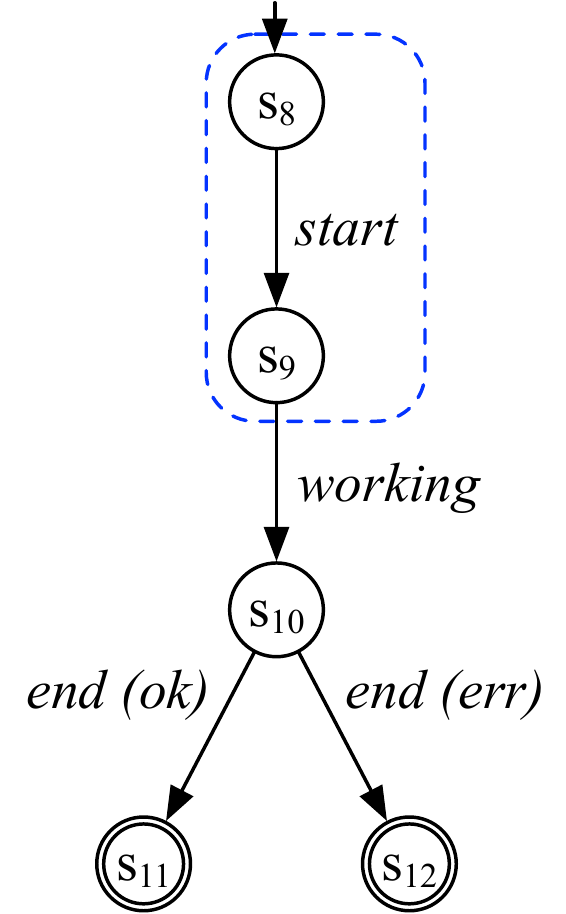}}
	\hfill
	\subfloat[$m_\texttt{J}$\label{fig:job}]{\includegraphics[height=4.5cm]{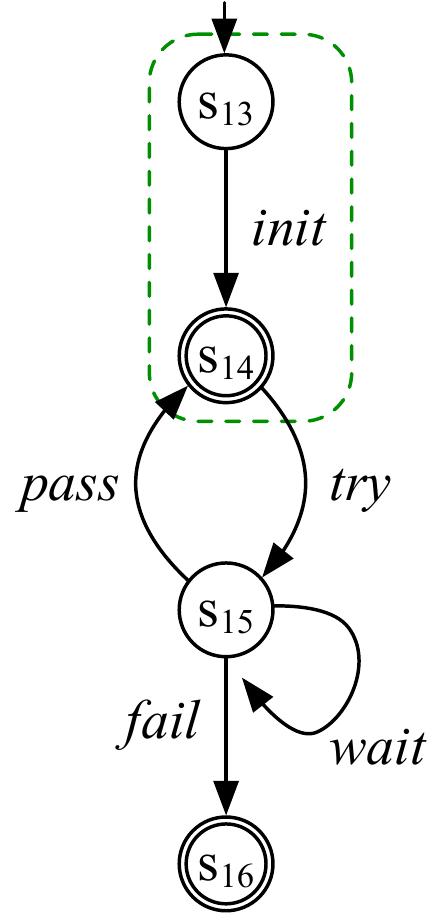}}
	\caption{Models corresponding to the running example logs ($m_\texttt{S}$: system model, $m_\texttt{M}$: model for component \texttt{Master}, $m_\texttt{J}$: model for component \texttt{Job})}
	\label{fig:models}
\end{figure}

A software engineer is tasked with building a finite-state model of the system
that accurately captures the behavior observed in the logs. However, the
engineer cannot rely on the system source code since it is not available. This
is the case, for example, where the system is mainly composed of heterogeneous,
3rd-party components for which neither the source code nor the documentation are
available~\cite{8811931,Palmer19,Rios20}. The only information about the system
to which engineers have access is represented by the execution logs
$L_\texttt{S}$. To perform this task, the engineer uses a tool implementing
one of the state-of-the-art \emph{model inference
techniques}~\cite{walkinshaw2016inferring,mariani2017gk,Emam2018Inf} proposed in
the literature; the tool takes as input the logs $L_\texttt{S}$ and returns
the system model $m_\texttt{S}$ shown in \figurename~\subref*{fig:system}.
Intuitively, we can see that $m_\texttt{S}$ properly reflects the flow of events
recorded in $L_\texttt{S}$. However, when the engineer tries to execute the
model inference tool on much larger logs of the same system, she observes that
the tool does not terminate within a practical time limit (e.g., one day).
Indeed, due to the intrinsic complexity of the model inference
problem~\cite{GOLD1967447}, the time complexity of state-of-the-art model
inference algorithms is polynomial~\cite{10.1007/BFb0054059,Emam2018Inf} in the
size of the input logs.

\begin{figure}
\small
\centering
\begin{tabular}{lll|lll}
\toprule
\multicolumn{3}{c|}{First execution log $l_1$} & \multicolumn{3}{c}{Second execution log $l_2$} \\
ID & Component & Event & ID & Component & Event \\
\midrule
$e_{1,1}$  & \texttt{Master}  & \textit{start}    & $e_{1,2}$  & \texttt{Master}  & \textit{start} \\
$e_{2,1}$  & \texttt{Job}     & \textit{init}     & $e_{2,2}$  & \texttt{Job}     & \textit{init} \\
$e_{3,1}$  & \texttt{Master}  & \textit{working}  & $e_{3,2}$  & \texttt{Master}  & \textit{working} \\
$e_{4,1}$  & \texttt{Job}     & \textit{try}      & $e_{4,2}$  & \texttt{Job}     & \textit{try}  \\
$e_{5,1}$  & \texttt{Job}     & \textit{pass}     & $e_{5,2}$  & \texttt{Job}     & \textit{wait} \\
$e_{6,1}$  & \texttt{Job}     & \textit{try}      & $e_{6,2}$  & \texttt{Job}     & \textit{wait} \\
$e_{7,1}$  & \texttt{Job}     & \textit{pass}     & $e_{7,2}$  & \texttt{Job}     & \textit{fail} \\
$e_{8,1}$  & \texttt{Master}  & \textit{end (ok)} & $e_{8,2}$  & \texttt{Master}  & \textit{end (err)} \\
\bottomrule
\end{tabular}
\caption{Running example logs $L_\texttt{S}=\{l_1, l_2\}$, inspired by Hadoop logs~\cite{he2020loghub}}
\label{fig:example-log}
\end{figure}

To address this problem, the engineer decides to use our new approach, \app: it
takes as input the logs $L_\texttt{S}$ in \figurename~\ref{fig:example-log} and
returns the \emph{same} system model $m_\texttt{S}$ shown in
\figurename~\subref*{fig:system}; the main difference with the tool used in the
previous attempts is that \app takes considerably less time to yield a system model.

The main idea behind \app is to tackle the intrinsic complexity of model
inference by means of a \emph{divide-and-conquer} approach: \app uses existing
model inference techniques to infer a model, not for the whole system but
\emph{for each component}. \figurename~\subref*{fig:master} and
\figurename~\subref*{fig:job} show the component models $m_\texttt{M}$ and
$m_\texttt{J}$ for the \texttt{Master} and \texttt{Job} components,
respectively. Component-level model inference is one of the main
contributors to the significant reduction of the execution time achieved by
\app. Furthermore, component-level model inference can be easily parallelized.

However, before yielding a system model, \app needs to properly ``combine'' the
individual component models. In our running example, this means building the
$m_\texttt{S}$ model shown in \figurename~\subref*{fig:system} by ``combining''
the component models in $M_C = \{m_\texttt{M}, m_\texttt{J} \}$ and shown in
\figurename~\subref*{fig:master} and \figurename~\subref*{fig:job}. This is a
challenging problem: we cannot simply concatenate or append the two component
models together, because the result would not conform to the flow of events
across the components recorded in the logs. In our running example logs, it is
recorded that the event \emph{start} of \texttt{Master} is immediately followed
by the event \emph{init} of \texttt{Job}. Such a flow of events recorded in the
logs should be represented in the final system model produced by \app. To
efficiently and effectively solve this problem, we propose a novel algorithm for
\emph{stitching} component models in the context of model inference.

\section{Scalable Model Inference}\label{sec:technique}

Our technique for scalable model inference follows a \emph{divide-and-conquer}
approach. The main idea is to first \emph{infer} a model of each system
component from the corresponding logs that are generated by the
\emph{projection} of system logs on the components; then, the individual
component models are \emph{stitched} together taking into account the
flows of the events across the components, \emph{as reflected in the logs}.
We call this approach \app (\emph{PR}ojection-\emph{IN}ference-\emph{S}titching).
The rationale behind \app is that, though existing (log-based) model inference
techniques cannot deal with the size of system logs, they can still be used to
accurately infer the models of individual components, since their logs are
sufficiently small for the existing model inference techniques to work. As
anticipated in Section~\ref{sec:example}, the challenge is then how to
``stitch'' together the models of the individual components to build a system
model that reflects not only the components' behaviors but also the flow of
events across the components, while preserving the accuracy of the component
models. Tackling this challenge is our main contribution, as detailed
in Section~\ref{sec:stitching}.

\begin{figure}
	\centering
	\includegraphics[width=0.9\linewidth]{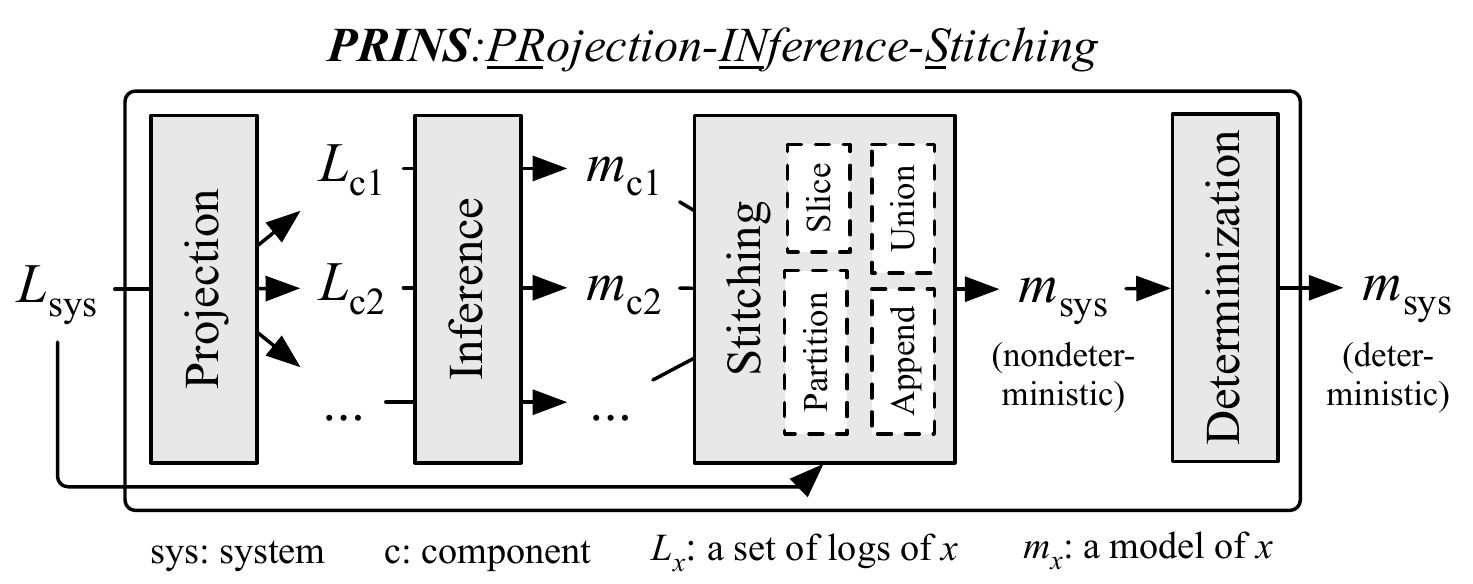}
	\caption{Overview of \app}
	\label{fig:overview}
\end{figure}

\figurename~\ref{fig:overview} outlines the workflow of \app. It takes as
input the logs of the system under analysis, possibly coming from multiple
executions; it returns a system model in the form of a gFSM. \app is composed
of four main stages: \emph{projection}, \emph{inference}, \emph{stitching}, and
\emph{determinization}.
The projection stage produces a set of logs for each component from the
input system logs. The component logs are then used to infer individual
component models in the inference stage. The stitching stage combines the
component models into a non-deterministic system model. Last, the
determinization stage transforms the non-deterministic model into a
deterministic model that is the output of \app.
The four stages are described in detail in the following subsections.

We remark that \app does not require any extra information (e.g., source
  code and documentation) other than logs. Furthermore, we do not
restrict logs to be produced by one thread/process: as one can see in the replication package for
our evaluation (see Section~\ref{sec:data}), individual logs for many of our
subject systems are already produced by multiple threads/processes
(distinguished by tid/pid).
Our only assumption is that log entries contain the name of
the ``component'' that generated them. This assumption is realistic
since this is common in practice, as shown
in real logs~\cite{he2020loghub}. Also, as indicated in
Section~\ref{sec:intro}, we use the term ``component'' in a broad sense (e.g.,
modules, classes) to represent an architectural ``part'' of a system. Therefore,
\app is applicable to any software system composed of multiple components as
long as their behavior is recorded in logs.

\subsection{Projection}\label{sec:projection}
This stage generates a set of component logs---which will be used to infer
a model for each component---from system logs. For instance, for our running example
logs $L_\texttt{S} = \{l_1, l_2\}$ shown in \figurename~\ref{fig:example-log},
we want to generate the set of logs for the \texttt{Master} component
$L_\texttt{M} =$ $\{\langle e_{1,1}, e_{3,1}, e_{8,1} \rangle$, $\langle e_{1,2},
e_{3,2}, e_{8,2}\rangle \}$, and the set of logs for the \texttt{Job}
component $L_\texttt{J} = \{\langle e_{2,1}, e_{4,1}, e_{5,1}, e_{6,1}, e_{7,1}
\rangle, \langle e_{2,2}, e_{4,2}, e_{5,2}, e_{6,2}, e_{7,2} \rangle
\}$. To achieve
this, we define the \emph{projection} operation as follows. Let $L$ be a set of
logs of a system and $C$ be a set of components of the system; the projection of
$L$ for a component $c\in C$, denoted with $L|c$, is the set of logs obtained from
$L$ by removing all occurrences of log entries of all $c'\in C$ where $c' \neq
c$. For the running example, we have $L_\texttt{S}|\texttt{Master} = L_\texttt{M}$
and $L_\texttt{S}|\texttt{Job} = L_\texttt{J}$.

\subsection{Inference}\label{sec:inference}
This stage infers individual component models from the sets of
component logs generated from the projection stage. This is
straightforward because inferring a (component) model from a set of
logs can be achieved using an off-the-shelf model inference
technique. We remark that \app does not depend on any particular model
inference technique, as long as it yields a deterministic FSM (or a
deterministic gFSM\footnote{A deterministic gFSM
  $m=(S, \mathit{ET}, G, \delta, s_0, F)$ with
  $\delta: S\times \mathit{ET}\times G\to S$ can be easily converted
  into a deterministic FSM $m'=(S, \Sigma, \delta', s_0, F)$ with
  $\delta': S\times \Sigma \to S$ where
  $\Sigma = \mathit{ET}\times G$.}) as a resulting model. Also, \app
can infer multiple component models in parallel because the inference
processes of the individual component models are independent from each
other. For the running example, using an off-the-shelf model inference
tool like MINT~\cite{minttool} on the logs in $L_\texttt{M}$ and
$L_\texttt{J}$, we obtain models $m_\texttt{M}$ (see
\figurename~\subref*{fig:master}) and $m_\texttt{J}$ (see
\figurename~\subref*{fig:job}), respectively.

We want to note that the parallelization of component model inference is just a  byproduct of using the divide-and-conquer approach enabled by the central component of \app: stitching (Section~\ref{sec:stitching}).

\subsection{Stitching}\label{sec:stitching}
Individual component models generated from the inference stage are
used in this \emph{stitching} stage, which is at the core of
\app. In this stage, we build a system model that captures not only
the components' behaviors inferred from the logs but also the flow of events
across components as reflected in the logs. For the running example, this means
building a model that is as ``similar'' as possible to $m_\texttt{S}$, using
models $m_\texttt{M}$ and $m_\texttt{J}$ as well as the input logs
$L_\texttt{S}$ reflecting the flow of events.

The idea of stitching comes from two important observations on the system
and component models:
\begin{compactenum}[(1)]
\item A system model is the composition of \emph{partial models}
  of the individual components; this means that \emph{partial behaviors}
  of components are \emph{interleaved} in a system model.
\item The component partial models (included within a system model)
  are combined together (i.e., appended) \emph{according to the
    flow of events recorded in logs}, since the system model must be
  able to accept the logs that were used to infer it.
\end{compactenum}
For example, in the models shown in \figurename~\ref{fig:models}, we can see
that the subgraph of $m_\texttt{S}$, enclosed with a red solid line,
contains two partial models: one, called $m_\texttt{M}^p$ and enclosed with a blue
dashed line, extracted from $m_\texttt{M}$ (including the states $s_8$ and
$s_9$---mapped to $s_0$ and $s_1$ in $m_\texttt{S}$---and the corresponding
transition labeled with \textit{start}) and the other, called
$m_\texttt{J}^p$ and enclosed with a green dashed line, extracted from
$m_\texttt{J}$ (including the states $s_{13}$ and $s_{14}$---mapped to $s_1$ and
$s_2$ in $m_\texttt{S}$---and the corresponding transition labeled with
\textit{init}). Notice that the partial models $m_\texttt{M}^p$ and
$m_\texttt{J}^p$ correspond to the partial (behaviors recorded in the) logs
$\langle e_{1,1} \rangle$ and $\langle e_{2,1} \rangle$, respectively, that are
determined by the interleaving of components in the system log $l_1$ shown in
\figurename~\ref{fig:example-log}. Furthermore, in $m_\texttt{S}$,
$m_\texttt{J}^p$ is appended to $m_\texttt{M}^p$, reflecting the fact that
event \textit{start} (from component \texttt{Master}) is immediately followed by
\textit{init} (from component \texttt{Job}) in the input logs.

Based on these observations, we propose a novel stitching technique
that first ``\emph{slices}'' individual component models into partial
models according to the component interleavings shown in logs; then it
``\emph{appends}'' partial models according to the flow of the events
recorded in logs. However, the behaviors of components recorded in
logs can be different from execution to execution (for instance, see
the difference in terms of recorded events between $l_1$ and $l_2$ in
our running example). To address this, we first build an intermediate, system-level
model \emph{for each execution} (i.e., for each log) and then merge
these models together at the end.

The \textit{\textbf{Stitch}} algorithm (whose pseudocode is shown in Algorithm~\ref{alg:stitch})
takes as input a set of logs $L_\mathit{sys}$ and a set of component models
$M_C$; it returns a system model $m_\mathit{sys}$ (built from the
elements in $M_C$) that accepts $L_\mathit{sys}$.

\begin{algorithm}
\SetKwInOut{Input}{Input}
\SetKwInOut{Output}{Output}

\Input{Set of System Logs (Structured) $L_\mathit{sys}$\\
Set of Component Models $M_C$}
\Output{System Model $m_\mathit{sys}$}

Set of Models $A \gets \emptyset$\\
\ForEach{Log $l_\mathit{sys} \in L_\mathit{sys}$}{\label{alg:st:beginL}
    Model $m_a \gets \mathit{emptyModel}$\label{alg:st:initm}\\
    $\mathit{initializeSliceStartStates}(M_C)$\label{alg:st:initSlice}\\
    List of Logs $P \gets \textit{\textbf{Partition}}(l_\mathit{sys})$\label{alg:st:partition}\\
    \ForEach{Log $l_c \in P$}{\label{alg:st:beginP}
        Model $m_c \gets \mathit{getCorrespondingModel}(c, M_C)$\\
        Model $m_\mathit{sl} \gets \textit{\textbf{Slice}}(m_c, l_c)$\label{alg:st:slice}\\
        $m_a \gets \textit{\textbf{Append}}(m_a, m_\mathit{sl})$\label{alg:st:append}\\
    }\label{alg:st:endP}
    $A.\mathit{add}(m_a)$\\
}\label{alg:st:endL}
Model $m_\mathit{sys} \gets \textit{\textbf{Union}}(A)$\label{alg:st:union}\\
\textbf{return} $m_\mathit{sys}$\label{alg:st:return}\\

\caption{\textit{\textbf{Stitch}}}
\label{alg:stitch}
\end{algorithm}

The algorithm builds a system-level model $m_a$ for each system log
$l_\mathit{sys} \in L_\mathit{sys}$
(lines~\ref{alg:st:beginL}--\ref{alg:st:endL}). To build $m_a$ for a given
$l_\mathit{sys}$, the algorithm first initializes $m_a$ as an empty model
(line~\ref{alg:st:initm}) and initializes the start states of all components
models in $M_C$ to their initial states (line~\ref{alg:st:initSlice}). The
algorithm then partitions $l_\mathit{sys}$ into a list of logs $P$, each one
corresponding to log entries of one component, according to the component
interleavings shown in $l_\mathit{sys}$ (using algorithm
\textit{\textbf{Partition}} at line~\ref{alg:st:partition}, described in detail
in Section~\ref{sec:partition}). For each log $l_c\in P$
(lines~\ref{alg:st:beginP}--\ref{alg:st:endP}), the algorithm retrieves the
component model $m_c \in M_C$ of the component $c$ that produced $l_c$, slices
it (using algorithm \textit{\textbf{Slice}} at line~\ref{alg:st:slice},
described in detail in Section~\ref{sec:slice}) into a partial model $m_\mathit{sl}$
that accepts \emph{only} log $l_c$, and then appends $m_\mathit{sl}$ to $m_a$
(using algorithm \textit{\textbf{Append}} at line~\ref{alg:st:append},
described in detail in Section~\ref{sec:append}). During the iteration over the
system logs in $L_\mathit{sys}$, the resulting system-level models $m_a$ are
collected in the set $A$. Last, the models in $A$ are combined into a single
model $m_\mathit{sys}$ (using algorithm \textit{\textbf{Union}} at
line~\ref{alg:st:union}, described in detail in Section~\ref{sec:union}). The
algorithm ends by returning $m_\mathit{sys}$ (line~\ref{alg:st:return}),
inferred from all logs in $L_\mathit{sys}$.

Before illustrating an example for \textit{\textbf{Stitch}}, let us first
present the details of the auxiliary algorithms \textit{\textbf{Partition}},
\textit{\textbf{Slice}}, \textit{\textbf{Append}}, and \textit{\textbf{Union}}.

\subsubsection{\textit{\textbf{Partition}}}\label{sec:partition}
This algorithm takes as input a system log $l$ (i.e., a sequence of log entries
from various components); it partitions $l$ into a sequence of logs $P$, where
each log $l_c \in P$ is the longest uninterrupted sequence of log entries
produced by the same component, and returns $P$. By doing this, we can divide a
system log into component-level logs, each of which represents the longest
uninterrupted partial behavior for a component,
while preserving the flow of events across components as recorded in the system log.

For instance, when the function takes as input the running example log $l_1$, it
returns $P = \langle l_{c,1}, l_{c,2}, \dots, l_{c,5} \rangle$ where
$l_{c,1}=\langle e_{1,1} \rangle$, $l_{c,2}=\langle e_{2,1} \rangle$,
$l_{c,3}=\langle e_{3,1} \rangle$, $l_{c,4}=\langle e_{4,1}, e_{5,1}, e_{6,1},
e_{7,1} \rangle$, and $l_{c,5}=\langle e_{8,1} \rangle$.

\subsubsection{\textit{\textbf{Slice}}}\label{sec:slice}
This algorithm (whose pseudocode is shown in
Algorithm~\ref{alg:slice}) takes as input a component model $m_c$ and
a component log $l_c$; it returns a new model $m_\mathit{sl}$, which
is the sliced version of $m_c$ and accepts only $l_c$.

\begin{algorithm}
\SetKwInOut{Input}{Input}
\SetKwInOut{Output}{Output}

\Input{Component Model $m_c$\\Component Log $l_c = \langle e_1, e_2, \dots \rangle$}
\Output{Component Model $m_\mathit{sl}$}

Model $m_\mathit{sl} \gets \texttt{emptyModel}$\\
State $s \gets \mathit{getSliceStartState}(m_c)$\label{alg:slice:init}\\
\ForEach{Log Entry $e \in l_c$}{\label{alg:slice:beginfor}
    Guarded Transition $\mathit{gt} \gets \mathit{getGT}(m_c, s, e)$\\
    $m_\mathit{sl} \gets \mathit{addGTAndStates}(m_\mathit{sl}, gt)$\\
    $s \gets \mathit{getTargetState}(\mathit{gt})$\\
}\label{alg:slice:endfor}
$\mathit{updateSliceStartState}(m_c, s)$\label{alg:slice:update}\\
\textbf{return} $m_\mathit{sl}$\\
\caption{\textit{\textbf{Slice}}}
\label{alg:slice}
\end{algorithm}

First, the algorithm retrieves the state of $m_c$ that will become the initial
state $s$ of the sliced model $m_\mathit{sl}$ (line~\ref{alg:slice:init}). Upon
the first invocation of \textit{\textbf{Slice}} for a certain model $m_c$, $s$
will be the initial state of $m_c$; for the subsequent invocations, $s$ will be
the last state visited in $m_c$ when running the previous slice operations. Note
that there is always only one last visited state because $m_c$ is deterministic,
as described in Section~\ref{sec:inference}. Starting from $s$, the algorithm
performs a run of $m_c$ as if it were to accept $l_c$ by iteratively reading
each log entry $e\in l_c$: the traversed states and guarded transitions of
$m_c$ are added into $m_\mathit{sl}$
(lines~\ref{alg:slice:beginfor}--\ref{alg:slice:endfor}). After the end of the
iteration, the algorithm records the last state $s$ visited in $m_c$
(line~\ref{alg:slice:update}), which is the (only one) final state of
$m_\mathit{sl}$ and will be used as the initial state of the next slice on
$m_c$. The algorithm ends by returning $m_\mathit{sl}$.

\begin{figure}
	\centering
	\includegraphics[width=0.8\linewidth]{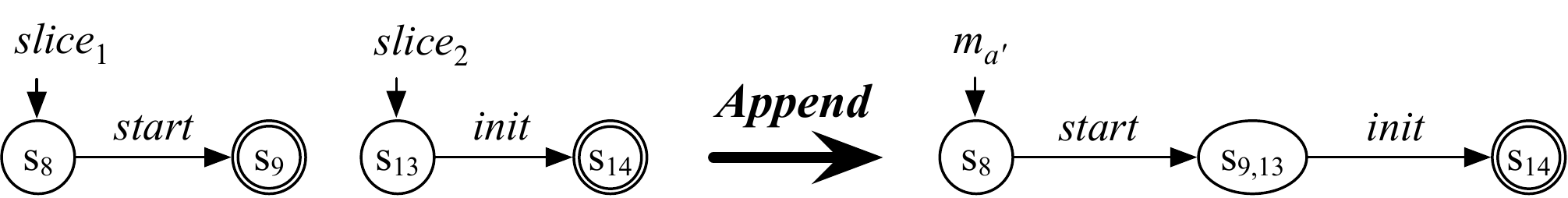}
	\caption{Illustration of appending two sliced models generated by \textit{\textbf{Slice}}
	for $l_{c,1}=\langle e_{1,1} \rangle$ and $l_{c,2}=\langle e_{2,1} \rangle$.
	They are appended by \textit{\textbf{Append}}, resulting in $m_{a'}$
	that accepts $\langle e_{1,1}, e_{2,1} \rangle$.}
    \label{fig:slice-append}
\end{figure}

For example, let us consider the case where \textit{\textbf{Slice}} is called
with parameters $m_c = m_\texttt{M}$ and $l_c = \langle e_{1,1} \rangle$, and
the slice start state returned by \textit{getSliceStartState} for $m_\texttt{M}$
is the initial state $s_8$. Starting from $s = s_8$, a run of $m_\texttt{M}$ is
performed: reading log entry $e_{1,1}$ results in making the guarded transition
to $s_1$ in $m_\texttt{M}$. This results in $m_\mathit{sl}$ to include the
guarded transition from $s_8$ to $s_9$ with label \textit{start} as well as the
states $s_8$ and $s_9$; the call to function \textit{getTargetState} updates $s$
to $s_9$. Since there is no more log entry in $l_c$, $s_9$ is the final state of
$m_\mathit{sl}$ and is the slice start state for the next call to
\textit{\textbf{Slice}} for $m_\texttt{M}$. The resulting $m_\mathit{sl}$ is
$\mathit{slice}_1$ shown in \figurename~\ref{fig:slice-append}.

\subsubsection{\textit{\textbf{Append}}}\label{sec:append}
This algorithm takes as input two models $m_a$ and $m_\mathit{sl}$; it returns
an updated version of $m_a$ built by appending $m_\mathit{sl}$ to the end of the
original version of $m_a$. If $m_a$ is an empty model (i.e., when
\textit{\textbf{Slice}} is called for the first time after the initialization of
$m_a$ in line~\ref{alg:st:initm} in Algorithm~\ref{alg:stitch}), the algorithm
simply returns $m_\mathit{sl}$. Otherwise, the algorithm merges the final state
of $m_a$ and the initial state of $m_\mathit{sl}$ and ends by returning the
updated $m_a$. Merging two states $s_x$ and $s_y$ is done by simply changing
both the source states of all outgoing transitions of $s_y$ and the target
states of all incoming transitions of $s_y$ as $s_x$. Note that a sliced model
$m_\mathit{sl}$ has only one final state as noted in Section~\ref{sec:slice}, and
therefore so does $m_a$.

For example, let us consider the case where \textit{\textbf{Append}} is called
with parameters $m_a = \mathit{slice}_1$ and $m_\mathit{sl} = \mathit{slice}_2$
shown in the left block of \figurename~\ref{fig:slice-append}.
The algorithm merges $s_9$ (i.e.,the final state of $m_a$) and $s_{13}$
(i.e., the initial state of $m_\mathit{sl}$), resulting in $m_{a'}$  shown
in the right block of \figurename~\ref{fig:slice-append}.

\subsubsection{\textit{\textbf{Union}}}\label{sec:union}
This algorithm takes as input a set of models $A$; it returns a model $m_u$ that
is able to accept all logs that can be accepted by all models in $A$. To do
this, the algorithm simply merges the initial states of all models in $A$, and
ends by returning the merged model as $m_u$.

We remark that merging states in \textit{\textbf{Append}} and
\textit{\textbf{Union}} can make the resulting model non-deterministic.
Actually, the two algorithms are simplified\footnote{To be precise, merging two
states is not equivalent to introducing an epsilon-transition from one to
another, but equivalent to introducing bi-directional epsilon-transitions
between the two states.} versions of the standard NFA (Non-deterministic Finite
Automata) concatenation and union operation, respectively. We will discuss
non-determinism later in the determinization stage (see Section~\ref{sec:det}).

\subsubsection{Application of \textit{\textbf{Stitch}} to the running example}
\begin{figure}
	\centering
	\includegraphics[width=0.9\linewidth]{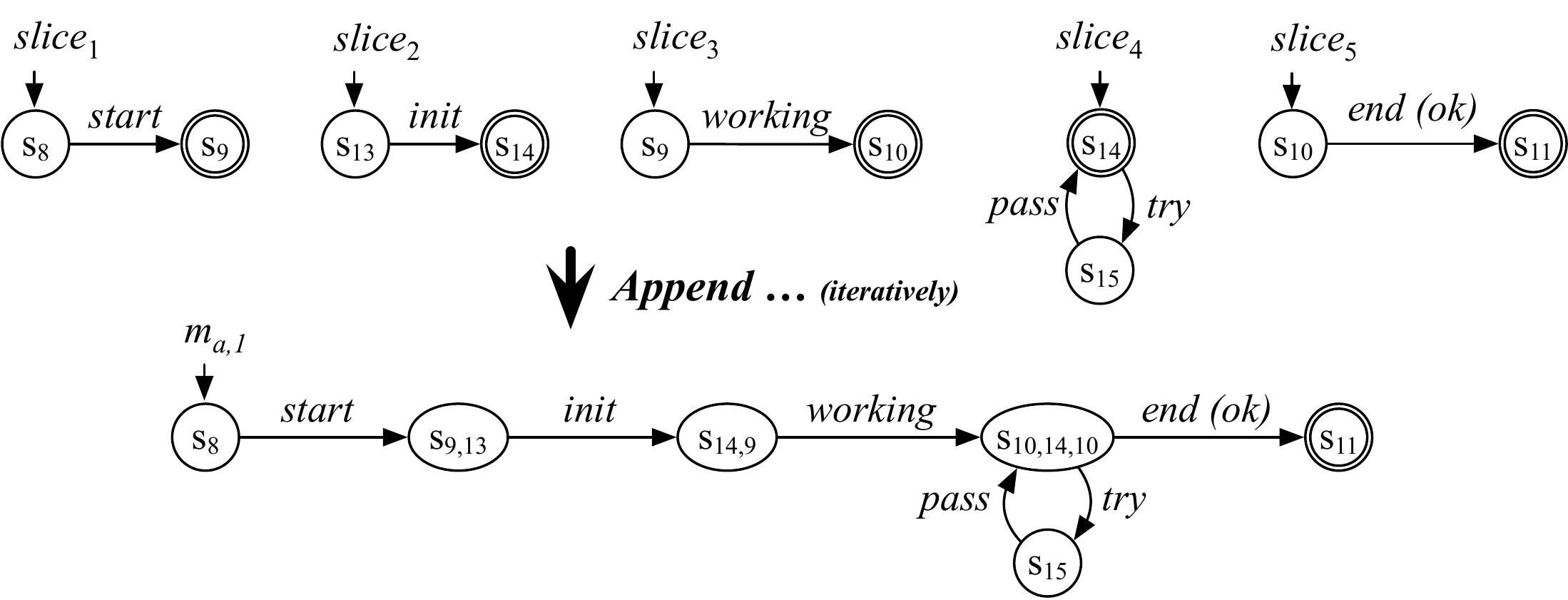}
	\caption{Illustration of building a system-level model for the running example
	log $l_1$. The five sliced models are generated by \textit{\textbf{Slice}}
	according to the partition of $l_1$. They are appended by
	\textit{\textbf{Append}} to build a system-level model $m_{a,1}$ that accepts $l_1$.}
    \label{fig:append}
\end{figure}

\begin{figure}
	\centering
	\includegraphics[width=0.9\linewidth]{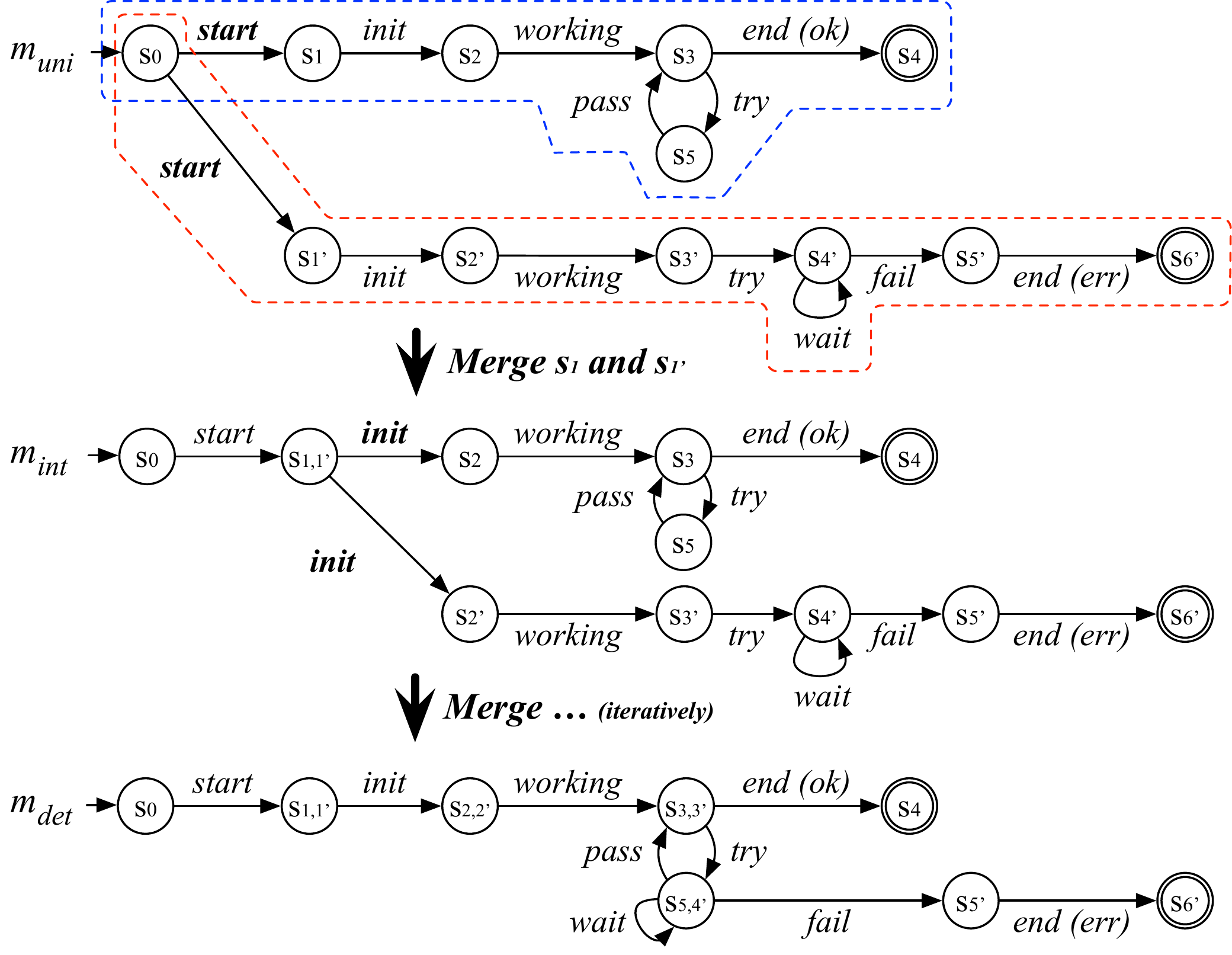}
	\caption{Illustration of the determinization using state merges for the system model
	build for the running example logs. The labels of nondeterministic transitions
	are highlighted in bold.}
	\label{fig:union-det}
\end{figure}

Let us consider the case where the \textit{\textbf{Stitch}} algorithm is called
with parameters $L_\mathit{sys}=\{l_1, l_2\}$ and $M_C=\{m_\texttt{M},
m_\texttt{J}\}$. For $l_1$, the call to \textit{\textbf{Partition}} yields $P =
\langle l_{c,1}, l_{c,2}, \dots, l_{c,5} \rangle$ where $l_{c,1}=\langle e_{1,1}
\rangle$, $l_{c,2}=\langle e_{2,1} \rangle$, $l_{c,3}=\langle e_{3,1} \rangle$,
$l_{c,4}=\langle e_{4,1}, e_{5,1}, e_{6,1}, e_{7,1} \rangle$, and
$l_{c,5}=\langle e_{8,1} \rangle$. For each $l_{c,i} \in P$, the call to
\textit{\textbf{Slice}} yields a sliced model $\mathit{slice}_i$ shown in the
top block of \figurename~\ref{fig:append}, originated from the component
models $m_\texttt{M}$ and $m_\texttt{J}$ shown in \figurename~\ref{fig:models}.
The five sliced models are appended to $m_a$ using \textit{\textbf{Append}},
resulting in a system-level model $m_{a,1}$ shown at the bottom of
\figurename~\ref{fig:append}. The state names of $m_{a,1}$ show how the
initial and final states of the sliced models were merged. For example,
$s_{10,14,10}$ is generated by merging $s_{10}$ (i.e., the final state of
$\mathit{slice}_3$), $s_{14}$ (i.e., the initial and final state of
$\mathit{slice}_4$), and $s_{10}$ (i.e., the initial state of $\mathit{slice}_5$).
Note that each $\mathit{slice}_i$ accepts the corresponding $l_{c,i} \in P$ and,
as a result, $m_{a,1}$ accepts $l_1$. The algorithm ends the iteration for $l_1$
by adding $m_{a,1}$ into $A$ and moves on to the next iteration to process
log $l_2$. After this second iteration completes, the newly built model $m_{a,2}$ for
$l_2$ is added to $A$; the call to \textit{\textbf{Union}} yields a system
model $m_\mathit{uni}$, shown at the top of \figurename~\ref{fig:union-det}. We
can see that $m_\mathit{uni}$ is composed of $m_{a,1}$ (i.e., the upper
subgraph enclosed with a blue dashed line) and $m_{a,2}$ (i.e., the lower
subgraph enclosed with a red dashed line).

In the above example, we can see that the output model $m_\mathit{uni}$ accepts
the input logs $L_\mathit{sys}$ as expected. However, $m_\mathit{uni}$ is
actually not equivalent to $m_\texttt{S}$ shown in
\figurename~\subref*{fig:system}: there exist potential logs that only $m_\texttt{S}$ can
accept, but $m_\mathit{uni}$ cannot. We will see how $m_\mathit{uni}$ can be
further transformed through the last stage of \app, i.e., determinization,
described in detail in Section~\ref{sec:det}.

\subsection{Determinization}\label{sec:det}

The last stage of \app post-processes the model yielded by the stitching stage
for mainly converting a non-deterministic model into a deterministic one.

Through the projection, inference, and stitching stages, we already get a system
model as an intermediate output. The non-determinism nature of such a model does
not represent an issue in many use cases (e.g., program
comprehension~\cite{Cook:1998:287001}, test case generation~\cite{6200086}).
However, especially when a model is used as an ``acceptor'' of (the behavior
recorded in) a log (e.g., in the case of anomaly detection~\cite{1541882}),
determinism is important for efficient checking. To broaden the use cases of
\app, we propose this determinization stage as an optional post-processing in \app.

The simplest way of converting a non-deterministic model into a deterministic
one is using standard algorithms, such as the powerset
construction~\cite{101145}, that guarantee the equivalence between the
non-deterministic model provided in input and the deterministic one returned as
output. However, the worst-case complexity of the powerset construction is
exponential in the size of the non-deterministic model, making it an
impracticable solution for many applications.

To tackle this issue, we introduce a new approach inspired by the
heuristic-based determinization approach proposed by~\citet{1566607}. Unlike the
powerset construction, their heuristic-based approach recursively
merges the target states of non-deterministic transitions starting from the
given state of the input model. While the idea of this approach is intuitive,
since it simply merges states during the process of determinization, it may
\emph{generalize} the model being determinized, meaning the determinized model
may accept additional logs that are not accepted by the original
non-deterministic model. Our preliminary evaluation found that this simple
strategy of merging states can produce an \emph{over}-generalized model by
merging too many states, especially when there are already many
non-deterministic transitions in the input model. To avoid such
over-generalization, we propose a new algorithm, called \emph{Hybrid
Determinization with parameter $u$} (HD$_u$), by combining ideas from the
heuristic-based determinization and the powerset construction methods.

Our HD$_u$ merges the target states of non-deterministic transitions, similar to
the heuristic-based determinization. However, to prevent over-generalization it
applies a heuristic: HD$_u$ does not merge a state with other
states\footnote{Regardless of the number of states to be merged, multiple states
can be merged into one state at once, not incrementally.} if the former has
already been merged $u$ times. The rationale behind this heuristic is to prevent
the merging of too many states, which causes the over-generalization. If
non-deterministic transitions remain because their target states are restricted
from being merged because of the value of $u$, HD$_u$ uses the powerset
construction to remove the remaining non-determinism while preserving the level
of the generalization. The larger the $u$ value is used, the more the model can
be generalized. The $u$ value can also be seen as the weight between the
heuristic-based determinization and the powerset construction; HD$_\infty$ is
the same as the heuristic-based determinization, while HD$_0$ is the same as the
powerset construction.

Algorithm~\ref{alg:hd} shows the pseudocode of HD$_u$. It takes as input a
non-deterministic model $m_n$ and a threshold $u$; it returns a deterministic
model $m_d$ that can accept all logs that can be accepted by $m_n$.

\begin{algorithm}
\SetKwInOut{Input}{Input}
\SetKwInOut{Output}{Output}

\Input{Model $m_n$\\Threshold $u$}
\Output{Model $m_d$}

Model $m_d \gets \mathit{copy}(m_n)$\\
Set of States $S_n \gets \textbf{\textit{getTargetStatesWithLimit}}(m_d, u)$\\
\While{$S_n \neq \emptyset$}{\label{alg:hd:beginIter}
    $\mathit{mergeStates}(m_d,S_n)$\\
    $S_n \gets \textbf{\textit{getTargetStatesWithLimit}}(m_d, u)$\\
}\label{alg:hd:endIter}
\If{$\mathit{isNonDeterministic}(m_d)$}{\label{alg:hd:beginStd}
    $m_d \gets \mathit{standardDeterminize}(m_d)$
}\label{alg:hd:endStd}
\textbf{return} $m_d$\\
\caption{Hybrid Determinization (HD$_u$)}
\label{alg:hd}
\end{algorithm}

The algorithm iteratively merges the set of states $S_n$ in $m_d$ as determined
by \textbf{\textit{getTargetStatesWithLimit}} (described below) until it is
empty (lines~\ref{alg:hd:beginIter}--\ref{alg:hd:endIter}). After the iteration
ends, if $m_d$ is still non-deterministic, the algorithm removes all the
remaining non-determinism using the powerset construction
(lines~\ref{alg:hd:beginStd}--\ref{alg:hd:endStd}). The algorithm ends by
returning $m_d$.

The heuristic to avoid the over-generalization is mainly implemented in function
\textbf{\textit{getTargetStatesWithLimit}} (whose pseudocode is shown in
Algorithm~\ref{alg:getnongt}). It takes as input a non-deterministic model $m_n$
and a threshold $u$; it returns a set of states $S_n$ to be merged to reduce
non-determinism in $m_n$, which does not contain the states that are restricted
from being merged because of the threshold $u$.

\begin{algorithm}
\SetKwInOut{Input}{Input}
\SetKwInOut{Output}{Output}

\Input{Model $m_n$\\Threshold $u$}
\Output{Set of States $S_n$}
\ForEach{Guarded Transition $\mathit{gt} \in getAllGT(m_n)$}{\label{alg:det:beginGT}
    Set of States $S_t \gets \mathit{getTargetStates}(\mathit{gt})$\label{alg:det:getT}\\
    Set of States $S_n \gets \mathit{removeAlreadyMergedStates}(S_t, u)$\label{alg:det:rm}\\
    \If{$|S_n| > 1$}{\label{alg:det:beginIf}
        \textbf{return} $S_n$\label{alg:det:return}\\
    }\label{alg:det:endIf}
}\label{alg:det:endGT}
\textbf{return} $\emptyset$\label{alg:det:empty}\\
\caption{\textbf{\textit{getTargetStatesWithLimit}}}
\label{alg:getnongt}
\end{algorithm}

For each guarded transition $\mathit{gt}$ in the transition relation of $m_n$
(lines~\ref{alg:det:beginGT}--\ref{alg:det:endGT}), the algorithm gets the set
of target states $S_t$ (line~\ref{alg:det:getT}), removes the states that have
already been merged $u$ times from $S_t$ to build $S_n$ (line~\ref{alg:det:rm}),
and returns $S_n$ (and ends) if it has more than one state
(lines~\ref{alg:det:beginIf}--\ref{alg:det:endIf}). If there is no such $S_n$
for all guarded transitions, the algorithm ends by returning $S_n = \emptyset$
(line~\ref{alg:det:empty}).

For example, let us consider the case where the
\textbf{\textit{getTargetStatesWithLimit}} algorithm begins the iteration for a
non-deterministic (guarded) transition whose target states are $s_\mathit{abc}$,
$s_d$, and $s_e$, with $u=1$. If $s_\mathit{abc}$ was generated by merging three
states $s_a$, $s_b$, and $s_c$, $S_t$ becomes $\{s_\mathit{abc}, s_d, s_e\}$ but
$S_n$ becomes $\{s_d, s_e\}$ because $\mathit{removeAlreadyMergedStates}$
excludes $s_\mathit{abc}$ (since it has been already merged once, given $u=1$).
Since $|S_n|=2$, the algorithm ends by returning $S_n = \{s_d, s_e\}$.

\figurename~\ref{fig:union-det} shows how HD works for our running example.
Recall that $m_\mathit{uni}$ is the intermediate output of the stitching stage.
Starting from the initial state, HD iteratively merges the target states of
non-deterministic transitions, such as $s_1$ and $s_1^\prime$ in
$m_\mathit{uni}$ and then $s_2$ and $s_2^\prime$ in $m_\mathit{int}$, until no
more non-deterministic transition remains; the resulting model is
$m_\mathit{det}$. We can see that $m_\mathit{det}$ is exactly the same as (i.e.,
is graph-isomorphic to) the ideal model $m_\texttt{S}$ in
\figurename~\ref{fig:models}.

Notice that HD$_u$ causes a reduction in
size of the models since it merges the target states of
non-deterministic transitions in the course of its heuristic
determinization. However, a more important question is to what extent the
accuracy of the resulting models varies because of size reduction. In
our empirical evaluation, we will assess the impact of using HD$_u$ on
the accuracy of the inferred models in \app with respect to the value of
threshold $u$. We will also investigate the execution time of HD$_u$
and devise practical guidelines for choosing the value of $u$ (see
Section~\ref{sec:eval}).

\section{Evaluation}\label{sec:eval}

In this section, we report on the evaluation of the performance of
\app in generating models of a component-based system from system logs.

First, we assess the execution time of \app in inferring models from
large execution logs. This is the primary dimension we focus on since
we propose \app as a viable alternative to state-of-the-art techniques
for processing large logs. Second, we analyze how accurate the models
generated by \app are. This is an important aspect because it is
orthogonal to scalability but has direct implications on the
feasibility of using the generated models to support software
engineering tasks (e.g., test case generation). However, the
execution time of \app and the accuracy of the models generated by
\app might depend on its configuration, i.e., the number of
parallel inference tasks in the inference stage (see Section~\ref{sec:inference})
and the parameter $u$ of HD$_u$ in the determinization stage
(see Section~\ref{sec:det}). Therefore, it is important to
investigate the best configurations of \app before comparing it to
state-of-the-art techniques.

Summing up, we investigate the following research questions:
\begin{enumerate}[\bf RQ1:]
\item \textit{How does the execution time of \app change according to the
	parallel inference tasks in the inference stage?}
\item \textit{How does the execution time of HD$_u$ change according to
	parameter $u$?}
\item \textit{How does the accuracy of the models (in the form of gFSMs)
	generated by HD$_u$ change according to parameter $u$?}
\item \textit{How fast is \app when compared to state-of-the-art model
    inference techniques?}
\item \textit{How accurate are the models
    generated by \app when compared to those generated by state-of-the-art
    model inference techniques?}
\end{enumerate}

\subsection{Benchmark and Settings}\label{sec:benchmark}

To evaluate \app, we assembled a benchmark composed of logs extracted
from two sources: the LogHub project~\cite{he2020loghub} and a
personal computer (PC) running desktop business applications on a
daily basis. Table~\ref{table:subjects} lists the systems we included
in the benchmark (grouped by source) and provides statistics about the
corresponding logs: the number of components (column \emph{\# Cmps}),
the number of logs\footnote{In LogHub~\cite{he2020loghub}, the
original dataset for HDFS contains \SI{575061} logs (distinguishable by
block-ids), but we used only 1000 logs that are randomly sampled from
all logs, since we found that the 1000 logs are representative enough,
as they contain all event templates that appear in all logs.} (column
\emph{\# Logs}), the number of event templates (column \emph{\#
Tpls}), and the total number of log entries\footnote{When additional
logging level information (e.g., \texttt{info}, \texttt{warn},
\texttt{debug}, \texttt{error}) was available for each log entry, we
only used the log entries of the two main levels, i.e., \texttt{info}
and \texttt{error}, as the others provide unnecessary details (e.g.,
the status of a specific internal variable) for building behavioral
models.} (column \emph{\# Entries}).

\begin{table}
\caption{Subject Systems and Logs}\label{table:subjects}
\centering
\begin{filecontents*}{benchmark.tex}
source,system,components,logs,templates,entries,confidence
LogHub,Hadoop,19,68,41,3575,0.999999997
LogHub,HDFS,8,1000,16,18741,0.954800911
LogHub,Linux,31,42,115,11259,0.781844212
LogHub,Spark,11,217,21,67725,0.984861254
LogHub,Zookeeper,18,36,40,25298,0.910055791
console,CoreSync,54,1418,204,30223,0.942047018
console,NGLClient,27,42,70,892,0.888468137
console,Oobelib,12,250,147,56557,0.889014845
console,PDApp,10,787,75,47394,0.873280811
\end{filecontents*}
\pgfplotstabletypeset[
	col sep=comma,
	every head row/.style={before row=\toprule,after row=\midrule},
	every last row/.style={after row=\bottomrule},
	every row no 5/.style={before row=\midrule},
    columns/source/.style={
    		column type=l,
    		column name=Source,
    		string type,
	    assign cell content/.code={%
			\ifnum\pgfplotstablerow=0
				\pgfkeyssetvalue{/pgfplots/table/@cell content}
					{\multirow{5}{*}{LogHub~\cite{he2020loghub}}}%
			\else
				\ifnum\pgfplotstablerow=5
					\pgfkeyssetvalue{/pgfplots/table/@cell content}
						{\multirow{4}{*}{PC}}%
				\else
					\pgfkeyssetvalue{/pgfplots/table/@cell content}{}%
				\fi
			\fi
		}
    },
    columns/system/.style={column type=l,column name=System,string type},
    columns/components/.style={column type=r,column name=\# Cmps},
    columns/logs/.style={column type=r,column name=\# Logs,1000 sep={\,},min exponent for 1000 sep=4},
    columns/templates/.style={column type=r,column name=\# Tpls},
    columns/entries/.style={column type=r,column name=\# Entries, 1000 sep={\,},min exponent for 1000 sep=4},
   	columns/confidence/.style={column type=r,column name=Conf,zerofill}
]{benchmark.tex}
\end{table}

LogHub~\cite{he2020loghub} is a data repository containing a large collection of
structured logs (and the corresponding event templates) from 16 different systems.
Among them, we selected the logs of the five systems based on two conditions:
\begin{inparaenum}[(1)]
\item the component (name or ID) for each log entry is available in the logs;
\item the number of logs for each system is more than 10.
\end{inparaenum}
We set condition \#1 because \app targets component-based systems; as
for condition \#2, we require a minimum number of logs to validate the
accuracy as part of RQ2 (see Section~\ref{sec:accr}).

To increase the diversity of our benchmark logs, we also included the
logs of a personal computer running daily for office use. We collected the
logs through the built-in \texttt{Console.app} application of macOS
10.15. Among the many logs available on the PC, we selected those
fulfilling the same two conditions stated above, ending up with four
systems. Additionally, to identify the events templates of the
unstructured log messages in these logs, we first used
state-of-the-art tools for log message format identification (i.e.,
Drain~\cite{8029742} and MoLFI~\cite{messaoudi2018search}) to compute
an initial set of templates and then manually refined them, e.g., by
collapsing similar templates into a single one. All the structured
logs (anonymized to hide sensitive information) are available online
(see Section~\ref{sec:data}).

To additionally evaluate whether the benchmark logs are
sufficient to infer models that faithfully represent actual system
behaviors, following another state-of-the-art model inference
study~\cite{Emam2018Inf}, we computed log confidence scores using the
formula provided by \citet{7371999}. Briefly speaking, a low
confidence score (e.g., $\leq 0.2$) indicates that the logs are not
sufficient, and therefore the model inferred from the logs are likely
not to be compatible with the actual behaviors of the system under
analysis. On the contrary, a high confidence score (e.g., $\geq 0.85$)
indicates that the logs are probably sufficient for the inferred model
to faithfully represent the actual system behaviors. Column
\emph{Conf} in Table~\ref{table:subjects} shows the confidence scores
calculated for our benchmark logs. The values suggest that the logs
are mostly sufficient to infer faithful models. Although the
confidence score for Linux (0.78) is lower than the other benchmarks
scores, the Linux logs are from an existing
benchmark~\cite{he2020loghub} and cannot therefore be improved.
Furthermore, since our main focus is to compare \app and other model
inference techniques using the same logs, having a somewhat moderate
confidence score is not a major threat to the validity of our
experiments.

We conducted our evaluation on a high-performance computing
platform\footnote{The experiments presented in this paper were
carried out using the HPC facilities of the University of
Luxembourg~\cite{VBCG_HPCS14} (see \url{https://hpc.uni.lu} for more details).},
using nodes equipped with Dell C6320 units (2 Xeon
E5-2680v4@\SI{2.4}{\giga\hertz}, \SI{128}{\giga\byte}). We allocated four
cores and \SI{16}{\giga\byte} per job.

\subsection{RQ1: Parallel Inference}\label{sec:parallel}
\subsubsection{Methodology}\label{sec:parallel:method}
To answer RQ1, we assessed the execution time of \app with different
parallelization configurations for its inference stage. Specifically,
we varied the maximum number of parallel workers (i.e., the maximum
number of parallel inference tasks) from one to four in steps of one to
investigate the relationship between the maximum number of parallel
workers and the execution time of \app. For example, when the number
is set to four, at most four workers are running in parallel to infer
four component models at the same time in the inference stage of \app.

To infer individual component models in the inference stage of \app,
we used MINT~\cite{walkinshaw2016inferring}, a state-of-the-art model
inference tool. We selected MINT because other tools are either not
publicly available or require additional information other than just
logs (e.g., source code or architectural design documents).
In all experiments, we used the same configuration of MINT
(i.e., minimum state merge score $k=2$ and AdaBoost as data classifier
algorithm), which we set based on the one used in a previous
study~\cite{walkinshaw2016inferring} conducted by the authors of MINT.

For each system in our benchmark, we ran the four configurations of
\app to infer a system model from the same logs and measured their
execution time. To account for the randomness in measuring execution
time, we repeated the experiment 10 times.

We remark that we disabled the determinization stage of \app because
it is not the main focus of RQ1. Determinization configurations will be
comprehensively investigated in RQ2 and RQ3.

\subsubsection{Results}\label{sec:parallel:results}

\begin{figure}
	\centering
	\includegraphics[width=\linewidth]{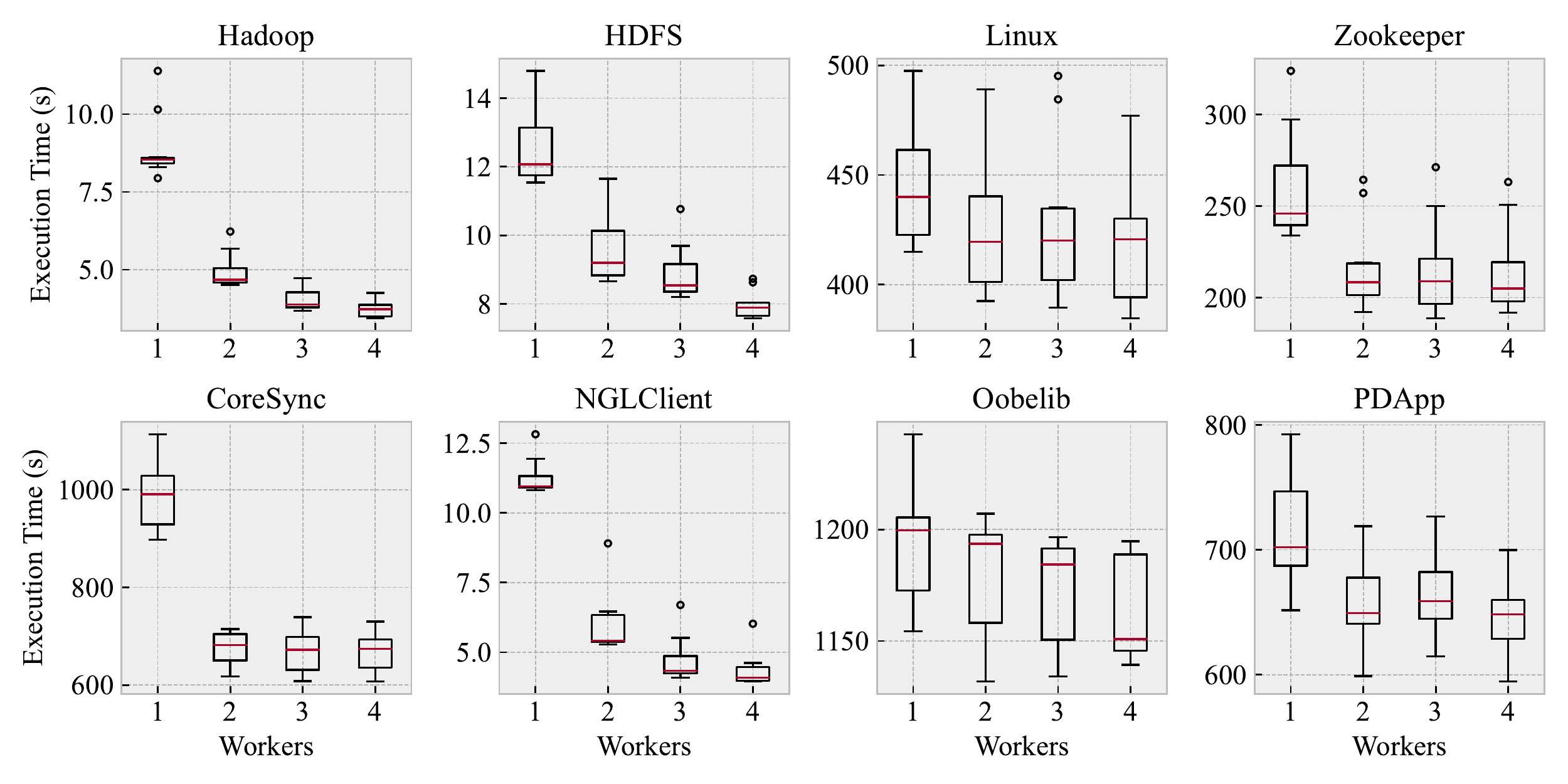}
	\caption{Relationship between the maximum number of parallel
          workers and the execution time of \app}
    \label{fig:parallel-boxplot}
\end{figure}

\figurename~\ref{fig:parallel-boxplot} shows the relationship between
the maximum number of parallel inference tasks (workers) and the execution time of \app.
None of the configurations was able to infer a model for Spark on all
ten executions due to out-of-memory errors. This occurred because
MINT (used by \app for inferring individual component models) could
not process the (huge) log of a component that is responsible for
producing about 97\% of all log messages of the system.

For all systems in our benchmark, it is clear that execution time
decreases as the maximum number of parallel inferences increases.
This is consistent with the general expectation for parallelization.

However, doubling the maximum number of parallel inferences does not
decrease the execution time in half. For example, there is  no clear
difference in execution time between two workers and four workers for
Linux, Zookeeper, CoreSync, and PDApp. A detailed analysis of the
results found that it is mainly because there are at most two major
components that take up more than 70\% of all log messages. For
example, Linux has two major components that represent around 50\% and
34\% of all log messages, while the third-largest component takes up
only 9.4\% of all messages. This implies that, for systems like Linux,
inferring component models is fast enough, except for a few major
components, and therefore having more than three parallel workers does
not significantly reduce execution time.

The answer to RQ1 is that the execution time of \app can be
significantly reduced by the parallel inference of individual
component models. However, the magnitude of the reduction in execution
time is not linear with respect to the maximum number of parallel inferences,
because not all components are equally sized in their logs. In
practice, an engineer can set the maximum number of parallel
inferences considering both available resources (e.g., the number of
CPUs and the total size of memory) and the log size distribution of
components.

\subsection{RQ2: Execution Time of Hybrid Determinization}\label{sec:det-time}
\subsubsection{Methodology}\label{sec:det-time:method}
To answer RQ2, we assessed the execution time of HD$_u$ with different
parameter values for $u$. Specifically, we varied the value of $u$
from one to ten in steps of one to investigate the relationship between the
value of $u$ and the execution time of HD$_u$. To additionally compare
HD$_u$ to the standard powerset construction, we also set $u=0$ (see
Section~\ref{sec:det} for more details).

For each system in our benchmark, we first ran \app without the
determinization stage to infer a non-deterministic system model. \app
internally used the same configuration of MINT as used in RQ1. For
each non-deterministic model, we ran HD$_u$ for all $u=0,1,\dots,10$ and
measured their execution time. To account for the randomness in
measuring execution time, we repeated the experiment 10 times.

\subsubsection{Results}\label{sec:det-time:results}

\begin{figure}
	\centering
	\includegraphics[width=\linewidth]{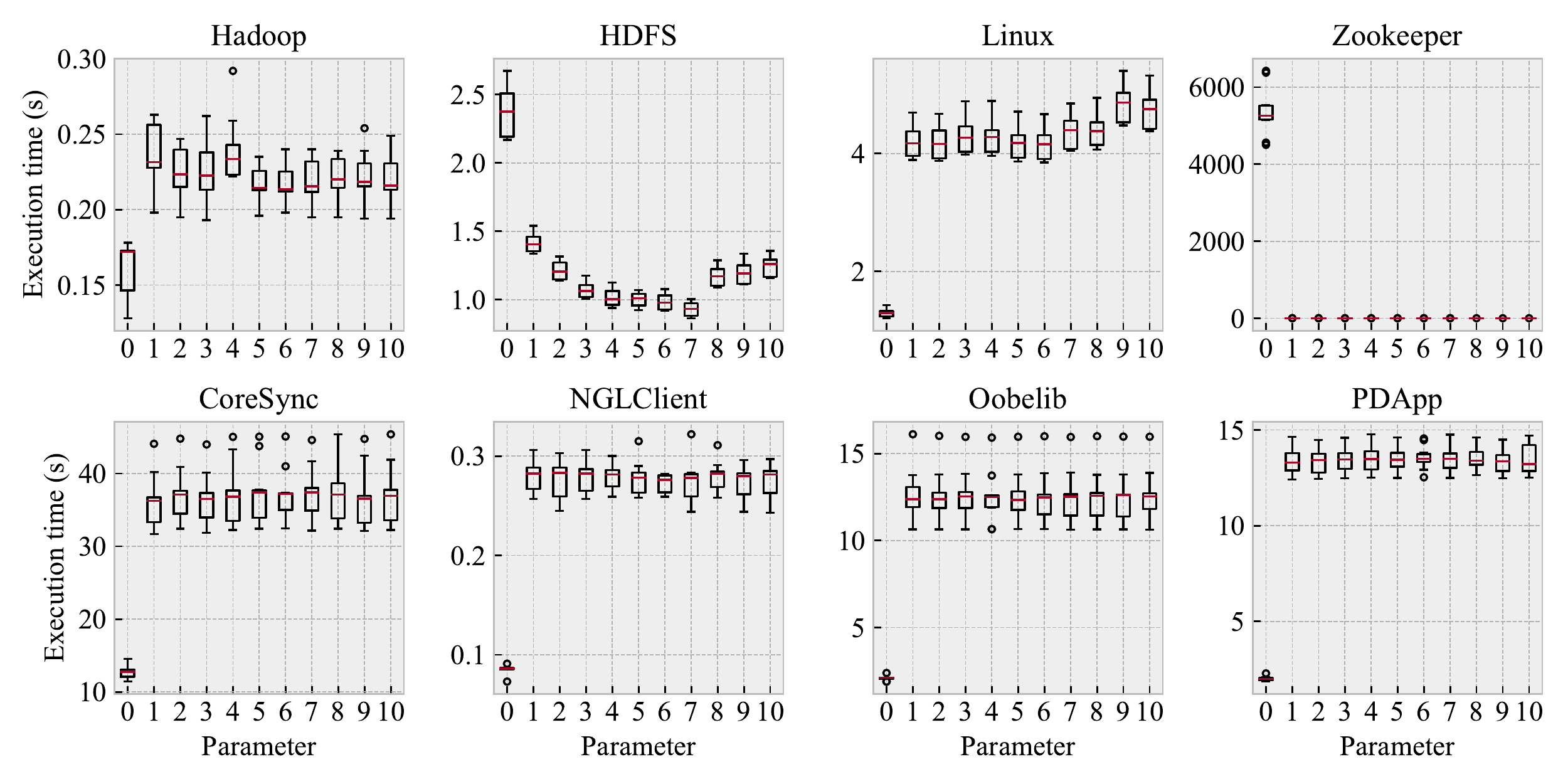}
	\caption{Relationship between the value of $u$ and the
          execution time of HD$_u$}
    \label{fig:det-time-boxplot}
\end{figure}

\figurename~\ref{fig:det-time-boxplot} shows the relationship between
the value of $u$ and the execution time of HD$_u$.  We have no results
for Spark because its non-deterministic system model was not available
for the reasons explained in Section~\ref{sec:parallel:results}.

For all the cases in which \app completed their execution for
generating non-deterministic system models, HD$_u$ (with $u \ge 1$)
took less than a minute. This implies that our hybrid determinization
can efficiently determinize non-deterministic models generated by
\app. On the other hand, the powerset construction (i.e., $u=0$) took
more than an hour for Zookeeper. This is due to the worst-case
complexity of the powerset construction as discussed in
Section~\ref{sec:det}. Interestingly, for the same non-deterministic
model of Zookeeper, using $u \ge 1$ significantly reduces the
determinization time. This clearly highlights the benefit of
hybrid determinization, i.e., combining the powerset construction and
the heuristic-based determinization.

The answer to RQ2 is that the execution time of HD$_u$ is practically
the same for all $u \ge 1$ because all of them are completed in less than
a minute. This implies that the best value of $u$ can be selected
mainly based on the accuracy of the resulting models, which will be
investigated in RQ3. On the other hand, the powerset construction
is indeed very time-consuming in extreme cases, which is consistent
with its well-known theoretical worst-case complexity. Since this cannot be predicted before running the determinization algorithms,
in practice, we can conclude that HD$_u$ with $u \ge 1$ is to be recommended over the standard powerset construction.

\subsection{RQ3: Accuracy of Models Generated by Hybrid Determinization}\label{sec:det-accr}
\subsubsection{Methodology}\label{sec:det-accr:method}
To answer RQ3, we first ran \app without the determinization stage to
infer a non-deterministic system model for each system in our
benchmark, as we did for RQ2. For each of the non-deterministic models, we then ran
HD$_u$ for all $u=0,1,\dots,10$ and measured the accuracy of the
deterministic models generated by HD$_u$.

We measured the accuracy in terms of \emph{recall},
\emph{specificity}, and Balanced Accuracy (BA), following previous
studies~\cite{1566607,walkinshaw2016inferring,
mariani2017gk,Emam2018Inf} in the area of model inference. Recall
measures the ability of the inferred models of a system to accept
``positive'' logs, i.e., logs containing feasible behaviors that the
system may exhibit. Specificity measures the ability of the inferred
models to reject ``negative'' logs, i.e., logs containing behaviors
that the system cannot exhibit. BA measures the balance between recall
and specificity and provides the summary of the two.

However, it is intrinsically difficult to evaluate the accuracy of inferred
models when there is no ground truth, i.e., reference models. To
address this issue,
we computed the metrics by using the well-known $k$-fold cross
validation (CV) method with $k=10$, which has also been used in
previous model inference studies~\cite{walkinshaw2016inferring,
mariani2017gk,Emam2018Inf}. This method
randomly partitions a set of logs into $k$ non-overlapping folds:
$k-1$ folds are used as ``training set'' from which the model inference
tool infers a model, while the remaining fold is used as
``test set'' to check whether the model inferred by the tool accepts
the logs in the fold. The procedure is repeated $k$ times until all
folds have been considered exactly once as the test set. For each
fold, if the inferred model successfully accepts a positive log in the
test set, the positive log is classified as True Positive (TP);
otherwise, the positive log is classified as False Negative (FN).
Similarly, if an inferred model successfully rejects a negative log in
the test set, the negative log is classified as True Negative (TN);
otherwise, the negative log is classified as False Positive (FP).
Based on the classification results, we calculated
$\mathit{recall}=\tfrac{|\mathit{TP}|}{|\mathit{TP}|+|\mathit{FN}|}$,
$\mathit{specificity}=\tfrac{|\mathit{TN}|}{|\mathit{TN}|+|\mathit{FP}|}$,
and the BA as the average of the recall and the specificity.

As done in previous work~\cite{walkinshaw2016inferring,
mariani2017gk,Emam2018Inf}, we synthesized negative logs from positive logs by
introducing small changes (mutations):
\begin{inparaenum}[(1)]
\item swapping two randomly selected log entries,
\item deleting a randomly selected log entry, and
\item adding a log entry randomly selected from other executions.
\end{inparaenum}
The changes should be small, because the larger the change is,
the easier an inferred model can detect the deviation of negative logs.
To further increase the probability\footnote{Recall that there are no reference models for the subject systems, and therefore we cannot verify if a synthesized log correctly contains an invalid system behavior.} that a log resulting from a mutation contains invalid behaviors of the
system, we checked whether the sequence of entries around the mutation location
(i.e., the mutated entries and the entries immediately before and after the
mutants) did not also appear in the positive logs.

\subsubsection{Results}\label{sec:det-accr:results}

\begin{figure}
	\centering
	\includegraphics[width=\linewidth]{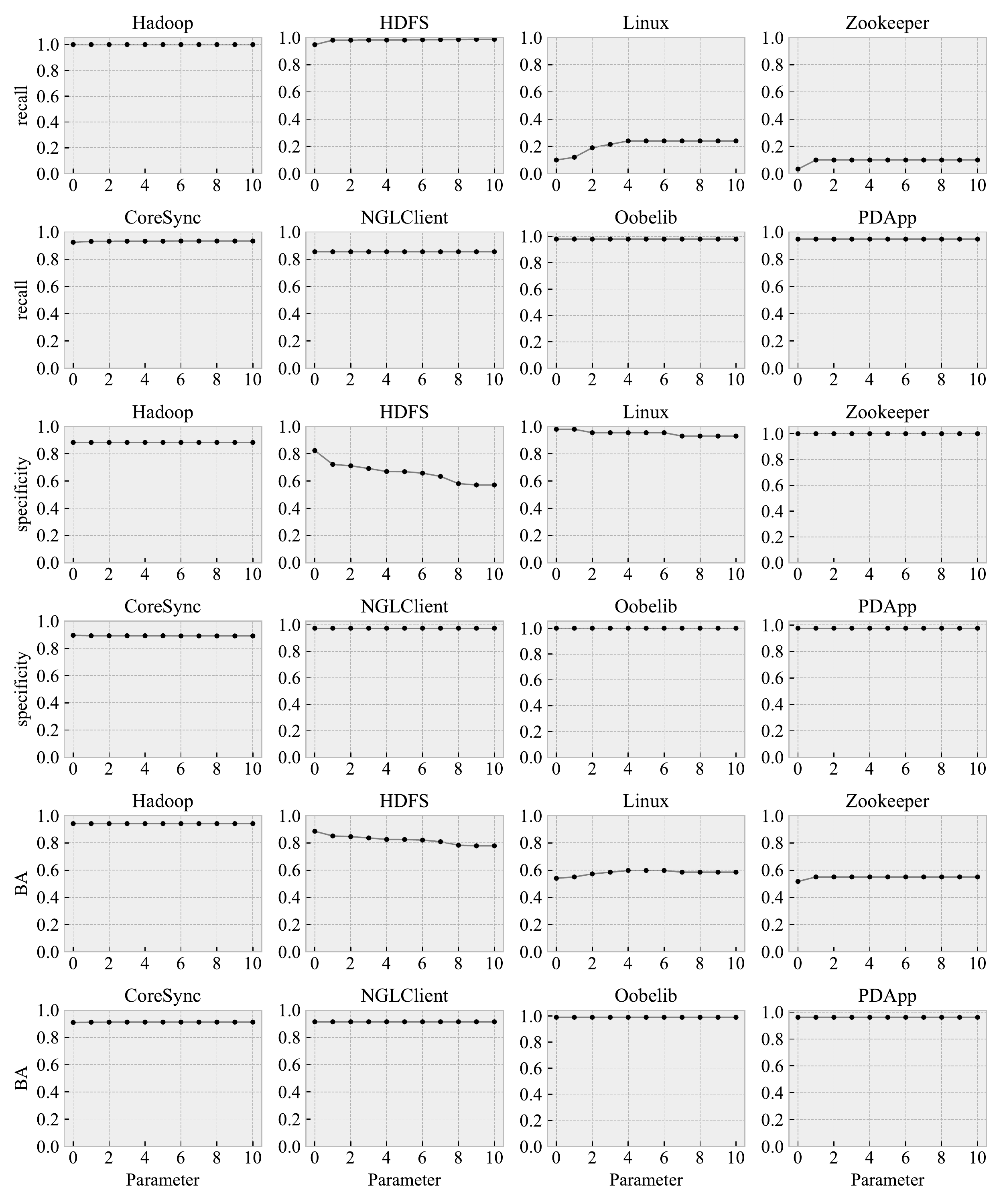}
	\caption{Relationship between the value of $u$ and the
          accuracy of the models generated by HD$_u$}
    \label{fig:det-accr-boxplot}
\end{figure}

\figurename~\ref{fig:det-accr-boxplot} shows the relationship between
the value of $u$ and the accuracy of the deterministic models generated by HD$_u$.
Again, Spark is not shown for the reasons explained in Section~\ref{sec:parallel:results}.

For Hadoop, CoreSync, NGLClient, Oobelib, and PDApp, there is no
change in recall, specificity, and BA when the value of $u$ changes.
This means that, for these five systems, the accuracy of deterministic
models generated by HD$_u$ does not change when the value of parameter
$u$ changes (for $u=0,1,\dots,10$). Furthermore, considering the fact that the powerset
construction (i.e., HD$_0$) guarantees the equivalence between the
non-deterministic model provided in input and the deterministic one
returned as output, identical accuracy for $u=0,1,\dots,10$ also implies that, regardless
of the value of $u$, HD$_u$ can convert non-deterministic models into
deterministic ones without sacrificing model accuracy for
five out of the eight systems in our benchmark, regardless
of the value of $u$.

For HDFS, Linux, and Zookeeper, as the value of $u$ increases,
recall values increase while specificity values decrease. This
means that, for the deterministic models generated by HD$_u$, if we
increase the value of $u$, then the ability to correctly accept
positive logs is improved whereas the ability to correctly reject
negative logs is diminished. This is intuitive because increasing the
value of $u$ merges more states to remove non-deterministic
transitions, yielding a generalized model that accepts more logs than
the original, non-deterministic model.

However, we can distinguish the increase in recall and the decrease in
specificity, because the former happens when the recall values of the input
non-deterministic models are close to zero (i.e., Linux
and Zookeeper), whereas the latter happens when the specificity values
of the non-deterministic
models are around 0.8. Since the non-deterministic
models of Linux and Zookeeper were already incapable of correctly
accepting positive logs, slightly improving them with determinization
is not practically significant. In fact, the logs of Linux and Zookeeper
were already inadequate for model inference in general, which will be
discussed in detail in Section~\ref{sec:accr:results}. On the other hand,
the decrease in specificity for HDFS is significant for HD$_u$ since it
should preserve the high specificity of the non-deterministic model
provided in input as much as possible. As a result, practically
speaking, the smaller the value of $u$, the better. Indeed, this
supports our idea that using $u$ to limit over-generalization in
hybrid determinization is helpful to avoid a significant accuracy loss.

The answer to RQ3 is that, for five out of eight systems in our
benchmark, the value of $u$ does not affect the accuracy of the
deterministic models generated by HD$_u$. However, for one system,
the accuracy practically decreases as the value of $n$ increases.
Additionally considering the high execution time of HD$_0$ (RQ2),
we can therefore conclude that $u=1$ is the best configuration trade-off for HD$_u$ in
terms of both execution time and accuracy in practice.

\subsection{RQ4: Execution Time of \app Compared to State-of-the-Art}\label{sec:time}
\subsubsection{Methodology}\label{sec:time:met}

To answer RQ4, we assessed the scalability of \app, in terms of
execution time, in comparison with MINT~\cite{walkinshaw2016inferring},
the same tool that is used internally by \app to generate
component-level models. In other words, we used
\emph{two} instances of MINT: the one used for the comparison in
inferring system models; the other one used internally by \app.
By doing this, we investigated to what extent the execution time of
model inference can be improved by using the divide-and-conquer
approach of \app compared to using a vanilla model inference.

Recall that \app can naturally infer many component models in
parallel at the inference stage, which can further improve the
execution time of \app as shown by the result of RQ1. To further investigate
this aspect, we used \emph{two} configurations of \app:
\app-\textit{P} where the parallel inference is enabled and
\app-\textit{N} where no parallelization is used. For \app-\textit{P},
we set the maximum number of parallel inferences to four, based on the
result of RQ1 and the number of allocated nodes (as described in
Section~\ref{sec:benchmark}). For both \app-\textit{P} and
\app-\textit{N}, we used the determinization stage (i.e., HD$_u$),
since MINT produces a deterministic model. For the value of $u$, we
used $u=1$ based on the results of RQ2 and RQ3.

We also varied the size of input logs to better understand the
impact of using larger logs on the execution time of \app and MINT. To
systematically increase such size while preserving the system behaviors
recorded in individual logs, we duplicated each of the logs following
the experiment design of \citet{7886964}. For example, when the
duplication factor is set to eight for the 250 logs (\num{56557} log entries)
of Oobelib, each of the 250 logs is duplicated eight times, and therefore
a total of $250 \times 8 = 2000$ logs 
($8 \times \num{56557} = \num{452456}$ log entries) are given as input both
to \app and to MINT. Notice that the system characteristics, such as the
number of components and the number of event templates, remain the
same when using duplicated logs. Since MINT could not infer models for large
logs due to out-of-memory failures or timeout (after 10 hours) in our preliminary
evaluation, we only varied the duplication factor from 1 to 8
in steps of 1.

For each set of duplicated logs for each system in our benchmark,
we ran MINT, \app-\textit{P}, and \app-\textit{N} to infer a
deterministic system model from the same logs and measured the
execution time of the tools.
To account for the randomness in measuring execution time, we repeated the
experiment three times and computed the average results.

\subsubsection{Results}\label{sec:time:results}

\begin{figure}
	\centering
	\includegraphics[width=\linewidth]{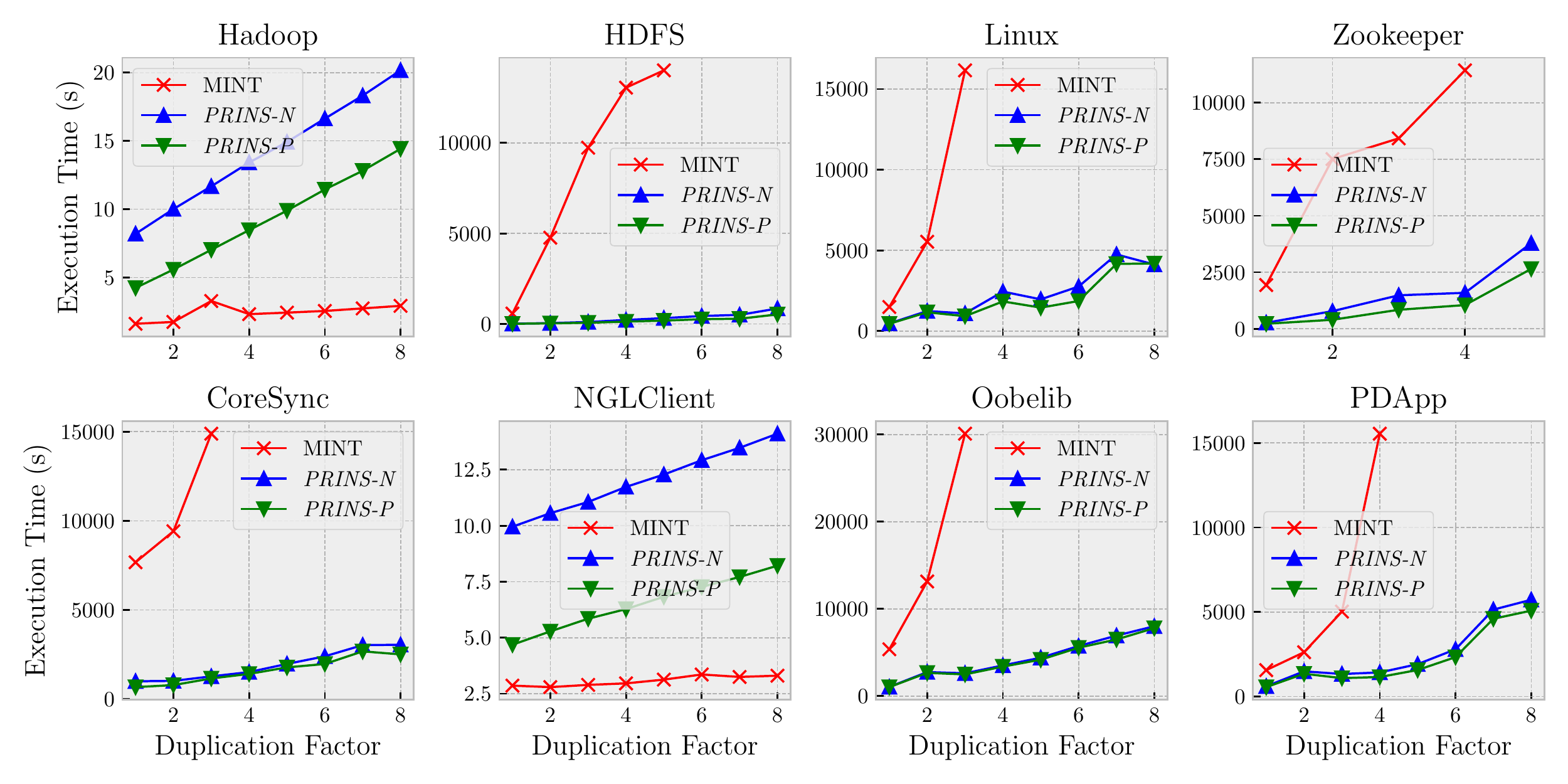}
	\caption{Comparison between MINT, \app-\textit{P}, and \app-\textit{N} in terms of execution time for various log sizes (obtained by varying the duplication factor of the benchmark logs)}
    \label{fig:mint-prins-time}
\end{figure}

\figurename~\ref{fig:mint-prins-time} shows the comparison
results between MINT, \app-\textit{P}, and \app-\textit{N}
in terms of execution time.
Because MINT (both the standalone instance and the one used
by \app) could not process the log of Spark, as explained in
Section~\ref{sec:parallel:results}, we have no results for it.
Also, due to the same reason, we have no results for
Zookeeper with a duplication factor above 5.

For all the cases in which at least one of the tools completed
their execution, we can see two distinct patterns in MINT's execution
time: for all of the duplicated logs, MINT completed its execution
only in two cases (Hadoop and NGLClient) out of eight; otherwise, MINT
could not complete its execution (HDFS, Linux, Zookeeper, CoreSync,
Oobelib, and PDApp).

However, for Hadoop and NGLClient, for which MINT
completed its execution for all of the duplicated logs, we can see
that MINT was quite fast (the execution time is less than \SI{5}{\s});
this can be attributed to the small size of the logs (\num{28600} entries
for Hadoop and \num{7136} entries for NGLClient even with a duplication
factor of 8). When the (standalone) MINT instance is already fast,
\app is actually slower than MINT due to the overhead for
projection, stitching, and determinization.
Nevertheless, even in these cases, using \app instead of MINT is still
practical since \app-\textit{P} took less than \SI{15}{\s}. Also, we
want to remark that such small logs are not really representative of
the large logs targeted by \app.

For the remaining six cases (HDFS, Linux, Zookeeper, CoreSync,
Oobelib, and PDApp) with larger log sizes, the
execution time of MINT increases steeply as the duplication factor
increases.  Furthermore, with a duplication factor above a
certain value (5 for HDFS, 4 for Zookeeper and PDApp, and 3 for Linux,
CoreSync, and Oobelib), MINT could not complete its execution due
to out-of-memory failures (Zookeeper) or timeouts after 10 hours (HDFS, Linux,
CoreSync, Oobelib, and PDApp). In contrast, for the same logs, the
execution time of \app-\textit{N} increases slowly as the duplication
factor increases, and there is no case where \app-\textit{N} could not
complete its execution (except for Zookeeper in which the MINT
instance internally used by \app for component model inference caused
an out-of-memory failure when the duplication factor is greater than 5). This
means that the scalability of model inference can be greatly improved
by using the divide-and-conquer approach of \app. Furthermore, for Zookeeper with a duplication factor
of 5, though MINT could not complete its execution due to out-of-memory failures, \app, on the other hand, completed successfully due to its divide-and-conquer approach.

Interestingly, for the six cases with larger log sizes, the difference between \app-\textit{P} and \app-\textit{N} is
very small compared to the difference between \app-\textit{N} and
MINT. This means that the key factor in the scalability improvement of
\app is the divide-and-conquer approach, not the parallel
inference of component models.

The answer to RQ4 is that the divide-and-conquer approach of \app
greatly improves the scalability of model inference for component-based
system logs and can even enable model inference when MINT leads to out-of-memory failures.

\subsection{RQ5: Accuracy of \app Compared to State-of-the-Art}\label{sec:accr}
\subsubsection{Methodology}\label{sec:accr:method}

To answer RQ2, we assessed the accuracy of the models inferred both by
\app and by MINT for each system in our benchmark, using the same
configuration for \app and MINT used as part of RQ4.
We measured the accuracy in terms of recall, specificity and BA as we did
for RQ3 (see Section~\ref{sec:det-accr:method} for more details).

\subsubsection{Results}\label{sec:accr:results}

\begin{table*}
\caption{Comparison between MINT (M) and \app (P) in terms of Accuracy. Differences between recall ($\Delta_R$), specificity ($\Delta_S$), and balanced accuracy ($\Delta_B$) values are expressed in percentage points ($\si{\pp}$); $\mathcal{LDS}$ is the log-component diversity score.}\label{table:mint-prins-accr}
\centering
\begin{filecontents*}{rq5-table.tex}
system,r_mint,r_prins,r_diff,s_mint,s_prins,s_diff,ba_mint,ba_prins,ba_diff,lcdiv
Hadoop,1,1,0,0.914285714,0.883333333,-3.095238095,0.957142857,0.941666667,-1.547619048,0.015
HDFS,0.981,0.98,-0.1,0.373,0.722,34.9,0.677,0.851,17.4,0.007
Linux,0.355,0.12,-23.5,0.885,0.98,9.5,0.62,0.55,-7,0.561
Zookeeper,0.216666667,0.1,-11.66666667,0.925,1,7.5,0.570833333,0.55,-2.083333333,0.571
CoreSync,0.947133154,0.930915992,-1.621716112,0.851243632,0.892767955,4.152432324,0.899188393,0.911841974,1.265358106,0.048
NGLClient,0.855,0.855,0,1,0.975,-2.5,0.9275,0.915,-1.25,0.195
Oobelib,0.98,0.98,0,1,1,0,0.99,0.99,0,0.016
PDApp,0.970723791,0.947874067,-2.284972412,0.977117819,0.975803311,-0.131450828,0.973920805,0.961838689,-1.20821162,0.014
Average,0.788190451,0.739223757,-4.896669399,0.865705896,0.928613075,6.290717925,0.826948174,0.833918416,0.697024263,0.178
\end{filecontents*}
\pgfplotstabletypeset[
	col sep=comma,
	every head row/.style={before row={
		\toprule
		& \multicolumn{3}{c}{Recall} & \multicolumn{3}{c}{Specificity} & \multicolumn{3}{c}{Balanced Accuracy}\\
		\cmidrule(r){2-4} \cmidrule(r){5-7} \cmidrule(r){8-10}
	},after row=\midrule},
	every even row/.style={before row={\rowcolor{lightgray}}},
	every last row/.style={before row=\midrule, after row=\bottomrule},
	fixed,fixed zerofill,precision=1,set thousands separator={},empty cells with={-},
    columns/system/.style={column type=l,column name=System,string type},
    columns/r_mint/.style={column type=r,column name=M,precision=2},
    columns/r_prins/.style={column type=r,column name=P,precision=2},
    columns/r_diff/.style={column type=r,column name=$\Delta_R$},
    columns/s_mint/.style={column type=r,column name=M,precision=2},
    columns/s_prins/.style={column type=r,column name=P,precision=2},
    columns/s_diff/.style={column type=r,column name=$\Delta_S$},
    columns/ba_mint/.style={column type=r,column name=M,precision=2},
    columns/ba_prins/.style={column type=r,column name=P,precision=2},
    columns/ba_diff/.style={column type=r,column name=$\Delta_B$},
    columns/lcdiv/.style={column type=r,column name=$\mathcal{LDS}$,precision=3,fixed}
]{rq5-table.tex}
\end{table*}

The accuracy scores of \app and MINT are shown in
Table~\ref{table:mint-prins-accr}. Under the \emph{Recall} column,
sub-columns \emph{M} and \emph{P} indicate the recall of MINT and
\app, respectively, and sub-column \emph{$\Delta_R$} indicates the
difference in recall between \app and MINT in percentage points
($\si{\pp}$). The sub-columns under the \emph{Specificity} and
\emph{Balanced Accuracy} columns follow the same structure, with
sub-column \emph{$\Delta_S$} indicating the difference in specificity
between \app and MINT, and sub-column \emph{$\Delta_B$} indicating the
difference in BA between \app and MINT. Again, none of the tools was
able to infer a model for Spark, for the reasons explained in
Section~\ref{sec:time:results}.

For all the cases in which the 10-fold CV completed without error,
the average difference in BA between \app and MINT is only
\SI[parse-numbers=false]{\avgBA}{\pp}, meaning that, on average, \app
is as accurate as MINT in inferring system models in terms of BA.
However, the average difference in recall between \app and MINT is
\SI[parse-numbers=false]{\avgRecall}{\pp}, while the average
difference in specificity between \app and MINT is
\SI[parse-numbers=false]{\avgSpec}{\pp}. This implies that, on average,
\app tends to infer models that are relatively less capable of
accepting positive logs but more capable of rejecting negative logs
than those inferred by MINT. The intuitive explanation is that a model
built by \app could be, in certain cases discussed below, more
specific to the flows of events recorded in individual input logs, due
to the way \app builds the model.
As described in Section~\ref{sec:stitching}, \app first builds an
intermediate system-level model \textit{for each execution log}
and then merges these intermediate models by merging only their
initial states at the end of the stitching. Though determinization
after stitching might further merge the other states for
removing non-determinism, it does so only for the states related to
non-deterministic transitions. Therefore, the execution-specific flows
of events captured in the intermediate system-level models can be
maintained (without being merged with the others) in the final system model
built by \app. In contrast, since MINT infers a model \textit{for all
system execution logs at once}, it tends to merge the
execution-specific flows of events to a larger extent than \app.
As expected, such characteristics also impact the size of inferred models. As
shown in \tablename~\ref{table:mint-prins-size}, the models inferred
by \app have on average 3.6 times more states and 5.5 times more
transitions than the models inferred by MINT. Since larger models are
more difficult to manually analyze and comprehend, this might be interpreted as a
drawback of \app. However, the models inferred by MINT are already too
large to be manually analyzed and understood, especially for systems with
large logs. Thus, automated techniques, such as model abstraction~\cite{4634784},
should be utilized in practice anyway.
Furthermore, inferred models can be used for other important
applications, such as test case generation~\cite{6200086} and anomaly
detection~\cite{1541882}, which do not require minimally sized models.
Therefore, the increased model size can be considered acceptable
given the significant scalability improvement reported in
Section~\ref{sec:time}.

\begin{table*}
\caption{Comparison between MINT and \app in terms of Model Size}\label{table:mint-prins-size}
\centering
\begin{filecontents*}{rq5-table-size.tex}
system,states_m,states_p,states_r,trans_m,trans_p,trans_r
Hadoop,67.0,65.0,0.9701492537313433,70.0,66.0,0.9428571428571428
HDFS,76.0,392.0,5.157894736842105,177.0,1308.0,7.389830508474576
Linux,342.0,1990.0,5.818713450292398,476.0,3322.0,6.9789915966386555
Zookeeper,376.0,3184.0,8.46808510638298,553.0,10667.0,19.289330922242314
CoreSync,3876.0,7524.0,1.9411764705882353,4318.0,10798.0,2.5006947660954144
NGLClient,148.0,154.0,1.0405405405405406,160.0,195.0,1.21875
Oobelib,447.0,1195.0,2.6733780760626398,545.0,1484.0,2.7229357798165137
PDApp,1301.0,3523.0,2.7079169869331285,1466.0,3801.0,2.592769440654843
Average,829.125,2253.375,3.597231827671671,970.625,3955.125,5.454520019597433
\end{filecontents*}
\pgfplotstabletypeset[
	col sep=comma,
	every head row/.style={before row={
		\toprule
		& \multicolumn{3}{c}{States} & \multicolumn{3}{c}{Transitions}\\
		\cmidrule(r){2-4} \cmidrule(r){5-7}
	},after row=\midrule},
	every even row/.style={before row={\rowcolor{lightgray}}},
	every last row/.style={before row=\midrule, after row=\bottomrule},
	fixed,fixed zerofill,precision=1,set thousands separator={},empty cells with={-},
    columns/system/.style={column type=l,column name=System,string type},
    columns/states_m/.style={column type=r,column name=MINT,precision=0},
    columns/states_p/.style={column type=r,column name=\app,precision=0},
    columns/states_r/.style={column type=r,column name=ratio},
    columns/trans_m/.style={column type=r,column name=MINT,precision=0},
    columns/trans_p/.style={column type=r,column name=\app,precision=0},
    columns/trans_r/.style={column type=r,column name=ratio}
]{rq5-table-size.tex}
\end{table*}

Looking at the results for individual systems, results differ
significantly in terms of $\Delta_R$ and $\Delta_S$ and it is
important to understand why to draw conclusions. For instance, for
HDFS, the value of $\Delta_S$ is high (\SI{34.9}{\pp}), while the value
of $\Delta_R$ is negligible. This shows that \app, compared to MINT,
can significantly increase the accuracy of the inferred models by
increasing their ability to correctly reject negative logs, without
compromising their ability to correctly accept positive logs.

On the other hand, for Linux and Zookeeper, the values of $\Delta_R$
are negative and practically significant (\SI{-23.5}{\pp} for Linux
and \SI{-11.7}{\pp} for Zookeeper) while the values of $\Delta_S$ are
positive and practically significant as well (\SI{34.9}{\pp} for Linux
and \SI{7.5}{\pp} for Zookeeper). Furthermore, the recall values of
both MINT and \app are relatively lower for Linux and Zookeeper
compared to the recall values for the other systems. In terms of the
10-fold CV, this means that the positive logs in the test set are not
properly accepted by the models inferred from the logs in the training
set for Linux and Zookeeper. Experimentally, this is mainly due to the
logs in the training set being \emph{too different} from the logs in
the test set, this being caused by the highly diverse logs of Linux
and Zookeeper overall. From a practical standpoint, this implies
that, regardless of the model inference technique, a model inferred
from existing logs may not be able to correctly accept unseen (but
positive) logs if the latter are too different from the former.
However, for the reasons mentioned above, the issue of highly diverse
logs has a moderately larger impact on \app than on MINT. Practical
implications are discussed below.

Before running model inference, to effectively predict and avoid cases
where \app is likely to be worse than MINT and where both techniques
fare poorly, we propose a new and practical metric to measure the
diversity of logs. Our log diversity metric is based on the
combination of components appearing in the individual logs because
\begin{inparaenum}[(1)]
\item \app targets component-based systems considering not only the
individual components' behaviors but also their interactions,
\item it is much simpler than using, for example, the flows of
log entries in the logs, and
\item it does not require any
extra information other than the logs.
\end{inparaenum}
More formally, let $L$ be a set of logs of a system and let $C(l)$ be
the set of components appearing in a log $l\in L$. We define
\emph{log-component diversity score} ($\mathcal{LDS}$) of the
system logs $L_\mathit{sys}$ as $\mathcal{LDS}(L_\mathit{sys}) =
\frac{U-1}{N-1}$, where $U = \lvert \{ C(l) \mid l\in L_\mathit{sys}
\} \rvert$ (i.e., the total number of unique $C(l)$s for all $l\in
L_\mathit{sys}$) and $N = \lvert L_\mathit{sys}\rvert$ (i.e., the
total number of logs in $L_\mathit{sys}$). In other words,
$\mathcal{LDS}$ indicates the ratio of logs that are unique
(i.e., different from the others) in terms of the set of components
appearing in the individual logs, ranging between 0 and 1; the higher
its value, the higher the diversity of the logs in terms of recording
different component interactions. For instance,
$\mathcal{LDS}(L_\texttt{S})=0$ for our running example logs
$L_\texttt{S} = \{l_1, l_2\}$ because $N=2$ and  $U= \lvert \{C(l_1),
C(l_2)\}\rvert = 1$  (since $C(l_1)$ $=$ $C(l_2)$ $=$
$\{\texttt{Master}, \texttt{Job}\}$). This means that $L_\texttt{S}$
is not diverse at all in terms of the appearing components.
Notice that $\mathcal{LDS}$ is a characteristic
of logs, which can be calculated before model inference takes place.

We measured $\mathcal{LDS}$ for the logs of each system in our
benchmark. Column $\mathcal{LDS}$ in
\tablename~\ref{table:mint-prins-accr} shows the results.
We can see that the resulting $\mathcal{LDS}$ values of Linux
(0.561) and Zookeeper (0.571) are much higher than those of the other
systems, which range between 0.007 (HDFS) and 0.195 (NGLClient). This
confirms that $\mathcal{LDS}$ can be effectively used to predict
whether the models inferred from the existing logs can correctly
accept unseen (but positive) logs or not before running model
inference.

In practice, if $\mathcal{LDS}$ is high (e.g., $> 0.2$)  for the
logs of a system, it implies that these logs do not sufficiently
exercise, in a comprehensive way, the potential behaviors of the
system. As a result, there is a high probability that many component
interactions have not been recorded or too rarely so. Therefore, an
engineer can address this problem by collecting more system logs until
$\mathcal{LDS}$ is low enough.

The answer to RQ5 is that, compared to MINT, \app tends to infer
models that are more capable of rejecting negative logs (i.e.,
yielding a higher specificity value) while sometimes being less
capable of accepting positive logs (i.e., yielding a lower recall
value). The latter happen anyway only in cases where logs are not
adequate for both techniques to work well. In practice, an engineer
can compute the diversity score of the logs before running model
inference, and easily determine whether more logs should be collected,
either through testing or usage, until the score is acceptable.

\subsection{Discussion and Threats to Validity}\label{sec:discussion}

From the results above, we conclude that \app is an order of magnitude
faster than MINT in model inference for component-based systems,
especially when the input system logs are large and the individual
component-level logs are considerably smaller than the system logs,
without significantly compromising the accuracy of the models.
Furthermore, since the large majority of modern software systems is
composed of many ``components'', which can be modules, classes,
or services, depending on the context, the logs typically encountered in
practice will satisfy the best conditions for \app to fare optimally:
the system logs are large but the individual component-level logs are
considerably smaller. There are situations where \app exhibits a
poorer recall than MINT. However, this is the case when the system
logs are inadequate for model inference in general, regardless of the
technique, and we have proposed a way to detect such situations and
remedy the problem.

One drawback of the divide-and-conquer approach in \app is the
increased size of inferred models. In this sense, \app can be seen as
sacrificing model size for improving the execution time of model
inference. Nevertheless, it is worth to note that \app does not
significantly compromise the accuracy of the inferred models.
Furthermore, given the significant execution time reduction in model
inference on large logs, increasing model size can be considered
acceptable.

In terms of threats to validity, using a specific model inference tool
(MINT) is a potential factor that may affect our results. However, we
expect that applying other model inference techniques would not change
the trends in results since the fundamental principles underlying the
different model inference techniques are very similar. Furthermore,
MINT is considered state-of-the-art among available tools.
Nevertheless, an experimental comparison across alternative
tools would be useful and is left for future work.

We used $k$-fold cross validation to evaluate the accuracy of
inferred models due to the lack of ground truth (i.e., reference
models) for our benchmark systems. Therefore, the computed accuracy
scores might not faithfully represent the similarity between the
inferred models and their (unknown) ground truths, especially when
the collected logs do not sufficiently represent the system behaviors.
To mitigate this issue, we calculated the log-confidence values,
following existing studies, and these results suggested that the logs
in our benchmarks are  sufficient to derive faithfully inferred models.
Furthermore, since the same logs are used for both \app and MINT,
not relying on ground truth does not severely affect our empirical
evaluation results.

\subsection{Data Availability}\label{sec:data}

The implementation of \app is available as a Python program.
The replication package,
including the benchmark logs and our implementation of \app, is
publicly available~\cite{shin_2022}. 
We also made a public GitHub repository at 
\url{https://github.com/SNTSVV/PRINS}.

\section{Related Work}\label{sec:related}

Starting from the seminal work of \citet{biermann1972synthesis} on the
\emph{k-Tail} algorithm, which is based on the concept of state
merging, several approaches have been proposed to infer a Finite State
Machine (FSM) from execution traces or logs.
\emph{Synoptic}~\cite{Beschastnikh2011Lev} uses temporal invariants,
mined from execution traces, to steer the FSM inference process to
find models that satisfy such invariants; the space of the possible
models is then explored using a combination of model refinement and
coarsening.
\emph{InvariMINT}~\cite{6951474} is an approach enabling the
declarative specification of model inference algorithms in terms of
the types of properties that will be enforced in the inferred model;
the empirical results show that the declarative approach outperforms
procedural implementations of \emph{k-Tail} and \emph{Synoptic}.
Nevertheless, this approach requires prior knowledge of the properties
that should hold on the inferred model; such a pre-condition cannot be
satisfied in contexts (like the one in which this work is set) where
the knowledge about the system is limited and the only information
about the system is provided by logs.
\emph{mk-Tails}~\cite{nimrod} is a generalization of the \emph{k-Tail}
algorithm  from single to many parameters, which enables fine-grained
control over the  abstraction (generalization) on different subsets of
the events. It allows users  to deal with the trade-off between size
and accuracy in model inference.

Other approaches infer other types of behavioral models that are
richer than an FSM. \emph{GK-tail+}~\cite{mariani2017gk} infers
guarded FSM (gFSM) by extending the \emph{k-Tail} algorithm and
combining it with Daikon~\cite{ERNST200735} to synthesize constraints
on parameter values; such constraints are represented as guards of the
transitions of the inferred model.
\emph{MINT}~\cite{walkinshaw2016inferring} also infers a gFSM by
combining EDSM (Evidence-Driven State Merging)~\cite{1600197} and data
classifier inference~\cite{witten2016data}. 
EDSM, based on the Blue-Fringe algorithm~\cite{10.1007/BFb0054059}, is a popular
and accurate model inference technique, which won the Abbadingo~\cite{10.1007/BFb0054059}
competition; it is also utilized in DFASAT~\cite{heule2013software} that won the
StaMinA competition~\cite{Walkinshaw2013}. Data-classifier inference
identifies patterns or rules between data values of an event and its
subsequent events. Using data classifiers, the data rules and their
subsequent events are explicitly tied together.
\emph{ReHMM} (Reinforcement learning-based Hidden Markov
Modeling)~\cite{Emam2018Inf} infers a gFSM extended with transition
probabilities, by using a hybrid technique that combines stochastic
modeling and reinforcement learning. ReHMM is built on top of MINT;
differently from the latter, it uses a specific data classifier
(Hidden Markov model) to deal with transition probabilities.

Model inference has also been proposed in the context of distributed
and concurrent systems. \emph{CSight}~\cite{Beschastnikh2014} infers
a communicating FSM from logs of vector-timestamped concurrent
executions, by mining temporal properties and refining the inferred
model in a way similar to \emph{Synoptic}.
\emph{MSGMiner}~\cite{Kumar2011} is a framework for mining graph-based
models (called Message Sequence Graphs) of distributed systems; the
nodes of this graph correspond to Message Sequence Chart, whereas the
edges are determined using automata learning techniques. This work has
been further extended~\cite{Kumar2012} to infer (symbolic) class level
specifications. However, these approaches require the availability of
channel definitions, i.e., which events are used to send and receive
messages among components.

Liu and Dongen~\cite{7849947} use a \emph{divide-and-conquer}
strategy, similar to the one in our \app approach, to infer a
system-level, hierarchical process model (in the form of a Petri net
with nested transitions) from the logs of interleaved components, by
leveraging the calling relation between the methods of different
components. This approach assumes the knowledge of the caller and
callee of each component methods; in our case, we do not have this
information and rely on the \emph{leads-to} relation among log
entries, computed from high-level architectural descriptions and
information about the communication events.

Nevertheless, all the aforementioned approaches cannot avoid
scalability issues due to the intrinsic computational complexity of
inferring FSM-like models; the minimal consistent FSM inference from
logs is NP-complete~\cite{GOLD1967447} and all the more practical
approaches are approximation algorithms with polynomial complexity.

One way to tackle the intrinsic scalability issue of (automata-based)
model inference is to rely on distributed computing models, such as
MapReduce~\cite{Dean2008}, by transforming the sequential model
inference algorithms into their corresponding distributed version. In
the case of the \emph{k-Tail} algorithm, the main
idea~\cite{wang2016scalable} is to parallelize the algorithm by
dividing the traces (sequences of log messages) into several groups,
and then run an instance of the sequential algorithm on each of them.
A more fine-grained version~\cite{LUO201713} parallelizes both the
trace slicing and the model synthesis steps. Being based on MapReduce,
both approaches require to encode the data to be exchanged between
mappers and reducers in the form of key-value pairs. This encoding,
especially in the trace slicing step, is application-specific;
for instance, to correctly slice traces recorded by an online
shopping system, different event parameter values, such as \texttt{user id},
\texttt{order id}, and \texttt{item id}, must be correctly identified and categorized
from individual messages beforehand. Notice that this is more
challenging than just identifying parameter values from free-formed
messages, since different types of parameters must be distinguished.
Hence, MapReduce cannot be used in contexts in which the system is
treated as a black-box, with limited information about the data
recorded in the log entries. Furthermore, though the approach can
infer a FSM from large logs of over 100 million events, the distributed
model synthesis can be significantly slower for $k \ge 3$
(of \emph{k-Tail}), since the underlying algorithm is exponential in $k$.

Another way of taming scalability is to
reduce the size of input logs by sampling them from the entire set of
collected logs using statistical analysis and provide statistical
guarantees on the inferred models. This is called \textit{statistical
log analysis} and was first presented by \citet{7886964}. Its key idea
is to iteratively sample new logs until the probability of adding new system
behaviors into the model inferred by sampled logs is less than a given
level of confidence threshold. While the idea of using statistical
analysis to address the scalability of model inference is promising,
as already noted by the authors, it is only applicable to
\textit{sequential} model inference algorithms, where each log can be
processed independently~\cite{7886964}. \app, on the other
hand, is applicable to all model inference algorithms as only the
inference target is changed from systems to components. Therefore, all
model inference algorithms can benefit from using the
divide-and-conquer approach in \app.

\section{Conclusion}\label{sec:conclusion}

In this paper, we addressed the scalability problem of inferring the
model of a component-based system from system logs, assuming that the
only information available about the system is represented by the
logs. Our approach, called \app, first infers a model of each system
component from the corresponding logs; then, it merges the individual
component models together taking into account the flow of events across
components, as reflected in the logs. Our evaluation, performed on
logs from nine datasets, has shown that \app can process large logs an
order of magnitude faster than a publicly available and well-known
state-of-the-art technique without significantly compromising the
accuracy of inferred models. While there are some cases where \app
achieves a moderately lower recall than the state-of-the-art, this
happens when the logs are inadequate for model inference in general,
regardless of the technique. Furthermore, we have proposed an easy way
to detect such cases and remedy the problem.

As part of future work, we plan to evaluate \app on different
datasets, especially collected from real-world industrial
applications, and to integrate it with other model inference
techniques. We also aim to assess the effectiveness of the inferred
models when applied to support software engineering activities,
such as test case generation.

\bibliographystyle{spbasic}
\bibliography{model-inference}

\end{document}